# The Correlation Conversion Property of Quantum Channels

Laszlo Gyongyosi

[1]Quantum Technologies Laboratory, Department of Telecommunications
*Budapest University of Technology and Economics*
2 Magyar tudosok krt, Budapest, H-1117, Hungary
[2]Information Systems Research Group, Mathematics and Natural Sciences
*Hungarian Academy of Sciences*
Budapest, H-1518, Hungary

gyongyosi@hit.bme.hu

**Abstract**

Transmission of quantum entanglement will play a crucial role in future networks and long-distance quantum communications. Quantum Key Distribution, the working mechanism of quantum repeaters and the various quantum communication protocols are all based on quantum entanglement. On the other hand, quantum entanglement is extremely fragile and sensitive to the noise of the communication channel over which it has been transmitted. To share entanglement between distant points, high fidelity quantum channels are needed. In practice, these communication links are noisy, which makes it impossible or extremely difficult and expensive to distribute entanglement. In this work we first show that quantum entanglement can be generated by a new idea, exploiting the most natural effect of the communication channels: the noise itself of the link. We prove that the noise transformation of quantum channels that are not able to transmit quantum entanglement can be used to generate distillable (useable) entanglement from classically correlated input. We call this new phenomenon the Correlation Conversion property (CC-property) of quantum channels. The proposed solution does not require any non-local operation or local measurement by the parties, only the use of standard quantum channels. Our results have implications and consequences for the future of quantum communications, and for global-scale quantum communication networks. The discovery also revealed that entanglement generation by local operations is possible.

One of the most important goals of current research in quantum computation and communications is the development of global-scale quantum communication networks. The success of worldwide Quantum Key Distribution and quantum repeater networks is based on quantum entanglement [1-7]. On the other hand, the process of entanglement sharing and distribution is an expensive task. The practical quantum channels are noisy, which makes it very hard or even impossible to send entangled particles over these links. The main reason is that quantum information is very fragile and extremely sensitive to the noise of the communication links. The current solutions under development for entanglement transmission are based on various entanglement purification methods, which could make it possible to share entanglement between distant points, but only if the noise of the communication links is low enough to allow the realization of the post-purification processes in the nodes. However, these purification methods are very expensive and inefficient, since many entangled pairs have to be shared between the parties with relatively high fidelity. One of the most fundamental questions in the development of future communication networks is the process of entanglement transmission. If it were possible to find quantum channels that could generate entanglement between two distant points (let us refer to them as Alice and Bob) without sending the entanglement itself, then we could dramatically reduce the cost of development of future quantum communication networks. It would also have very serious consequences for current knowledge about the nature of the information itself.

Over a quantum channel $\mathcal{N}$, many types of information can be transmitted. Sending entanglement would be possible only if the noise of channel $\mathcal{N}$ is low (i.e., it is a *high fidelity* channel which can transmit quantum information). On the other hand, if the noise of $\mathcal{N}$ is high (assuming it is a practical communication channel) then entanglement might be transmitted with much difficulty, or it could be completely impossible.

Let us assume that there is a quantum channel $\mathcal{N}$ between Alice and Bob, which is so noisy that it cannot function as a transmission venue for any quantum information, but it can be used to send classical information over it (i.e., it has quantum capacity $Q\left(\mathcal{N}\right) = 0$, but has positive classical capacity $C\left(\mathcal{N}\right) > 0$).

For simplicity, we will refer to this *quantum* channel $\mathcal{N}$ as a *classical-quantum channel* * (or *low fidelity* channel), since it can transmit classical correlations only. The condition $Q(\mathcal{N}) = 0$ also trivially holds for an entanglement-binding channel [3] that can produce only *non-distillable* (i.e., *useless* for quantum communication) bound entanglement. If Alice would like to send *distillable* (i.e., *useable* for quantum communication) *entanglement* to Bob over channel $\mathcal{N}$, she will find that it is not possible, since the noise of $\mathcal{N}$ makes it impossible to preserve the quantum entanglement. Alice must choose a different solution.

Since the channel between Alice and Bob is so noisy and the transmission of quantum entanglement is a difficult task, she might think the following: "*Since the channel is noisy and quantum entanglement is very fragile, would it be possible to feed only classical correlations to classical-quantum channel $\mathcal{N}$, to get quantum entanglement between my system, A, and system B on Bob's side?*"

In that case the problem of entanglement sharing would be reduced to the following process: Alice prepares a classically correlated system, *AB*, in which she keeps *A* and feeds *B* to channel $\mathcal{N}$. Bob receives *B*, and the result is quantum entanglement between Alice and Bob, generated simply by the noise of the channel.

If it were possible, Alice could use the classical-quantum channel $\mathcal{N}$ to send entanglement to Bob, except for the fact that she prepared a classically correlated input and the channel can transmit classical correlation only. *This idea might seem to be unimaginable and completely impossible at first sight, and our intuitions also strictly dictate that it cannot be true.*

Up to this point, the possibility of entanglement generation between Alice and Bob has been based on the transmission of quantum entanglement, which requires high fidelity quantum channels between the parties.

---

* *The term „classical-quantum channel" has several different interpretations in the literature. It is used mainly in the HSW (Holevo-Schumacher-Westmoreland) setting to describe the transmission of classical information over quantum channels. However, from the „classical-quantum" term it does not follow unambiguously that the quantum channel could not transmit quantum correlations. In our setting, under the „classical-quantum channel" we mean only those quantum channels that can transmit classical correlations, only.*

*As we have found, this is not the case.* There exist low fidelity channels which can transmit only classical correlation, but the noise transformation of the channel can re-transform the input density matrix in such a way that it will result in quantum entanglement between Alice and Bob. From this point onward, Alice has a much better choice than to send the entanglement directly over $\mathcal{N}$ . Alice can feed only a classically correlated input system to $\mathcal{N}$ , and the process of entanglement transmission will be made by the most natural property of these communication channels: *by the noise transformation of the channel,* itself. We called this new phenomenon the "*Correlation Conversion*" *property* (CC-property) of quantum channels.

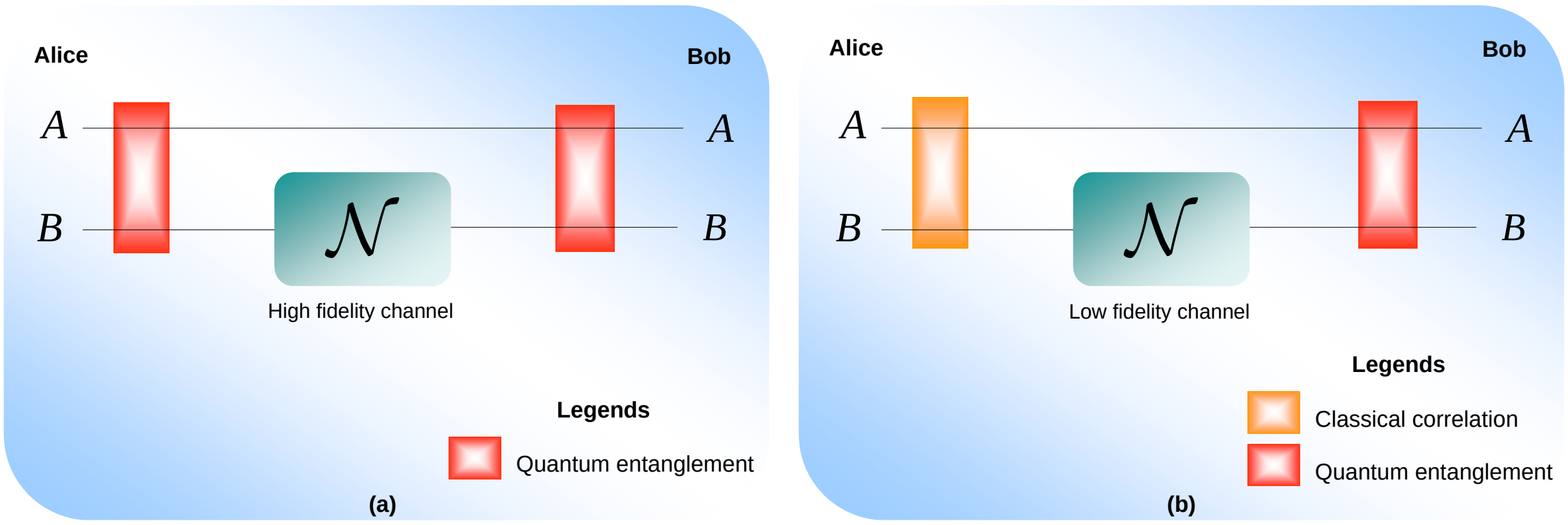

**Fig. 1.** (**a**): Alice's standard solution for sending entanglement to Bob. If she would like to send part $B$ of her entangled system $AB$ to Bob, then the channel has to be a high fidelity link, since quantum entanglement is extremely fragile and sensitive to noise. If the channel is noisy, the transmission of entanglement is a difficult task or completely impossible. (**b**): In our solution, Alice feeds only a classically correlated input to the classical (low fidelity) channel $\mathcal{N}$, which will result in quantum entanglement between her system $A$ and Bob's system $B$. The process does not require high fidelity channels, since the entanglement is generated by the noise transformation of the channel. This property is called the Correlation Conversion property of the communication link.

Producing entanglement from classical correlation by the noise of quantum channels seemed to be impossible before our results. However, it has already been shown that separable states can be used to distribute entanglement [8-12], but these protocols require ideal or nearly noiseless channels between Alice and Bob, which is completely unattainable in a practical communication sys-tem. These schemes also have another drawback: the requirement of non-local operations and local measurement. Our solution does not require ideal channels nor any non-local operation or local measurement on the encoder or decoder side to generate distillable quantum entanglement, only the use of standard quantum channels, i.e., *local operations* [15].

*The Correlation Conversion property of quantum channels is summarized as follows. There exist channels* $\mathcal{N}_1$ *and* $\mathcal{N}_2$ *which can produce quantum entanglement from classical correlated inputs* $\rho_{AB}$ *and* $\rho_{AC}$ *, between systems* $\rho_A$ *and channel output* $\sigma_B = \mathcal{N}_1(\rho_B)$*, where neither channel* $\mathcal{N}_1$ *nor* $\mathcal{N}_2$ *can transmit any quantum entanglement,* $Q(\mathcal{N}_1) = Q(\mathcal{N}_2) = 0$ *. The noise transformation of the channel can re-transform the density matrix in such a way that it results in entanglement between systems A and B.*

The channel construction $\mathcal{N}_1 \otimes \mathcal{N}_2$ is summarized in Fig. 2. Neither channel $\mathcal{N}_1$ nor $\mathcal{N}_2$ can transmit any quantum information or entanglement. The two channels can transmit classical correlation only. In the initial phase, Alice prepares the classically correlated systems $\rho_{AB}$ and $\rho_{AC}$. The input density matrices $\rho_A$ and $\rho_B$ are classically correlated and contain no quantum entanglement. Alice then feeds $\rho_B$ to the input of $\mathcal{N}_1$, and a *flag system* $\rho_C$, to the input of $\mathcal{N}_2$. The distillable quantum entanglement will be generated by the output of $\mathcal{N}_1(\rho_B)$, between density matrices $\rho_A$ and $\sigma_B = \mathcal{N}_1(\rho_B)$. The output $\sigma_C = \mathcal{N}_2(\rho_C)$ of the second channel will be also received by Bob, and will be simply traced out in the calculations. The final system state will be referred to as $\sigma_{AB} = Tr_C(\rho_A \mathcal{N}_1(\rho_B) \otimes \mathcal{N}_2(\rho_C))$, in which the system state will contain quantum entanglement between $\rho_A$ and $\sigma_B$.

We can easily find such kinds of channels; for example, any channel $\mathcal{N}_1$ with error probability $p = p_x + p_y + p_z$, could handle quantum entanglement transmission, but only if

$$Q(\mathcal{N}_1) = 1 - 2\left(p_x + p_y + p_z + \sqrt{p_x}\sqrt{p_y} + \sqrt{p_x}\sqrt{p_z} + \sqrt{p_y}\sqrt{p_z}\right) > 0 \quad (1)$$

holds true [13]. The error probability $p$ of $\mathcal{N}_1$ is so high that it makes it impossible to transmit quantum entanglement, thus $Q(\mathcal{N}_1) = 0$. The noise parameters $p_x, p_y$ and $p_z$ affect the eigen-values $v_+, v_-$ of the input density matrix of $\rho_{AB}$ as will be proven in the Supplementary Material. For the second channel, $\mathcal{N}_2$, the condition $Q(\mathcal{N}_2) = 0$ also has to hold. For an *entanglement-breaking* channel $\mathcal{N}_2$ this condition is trivially satisfied since it destroys every quantum entanglement on its output [17].

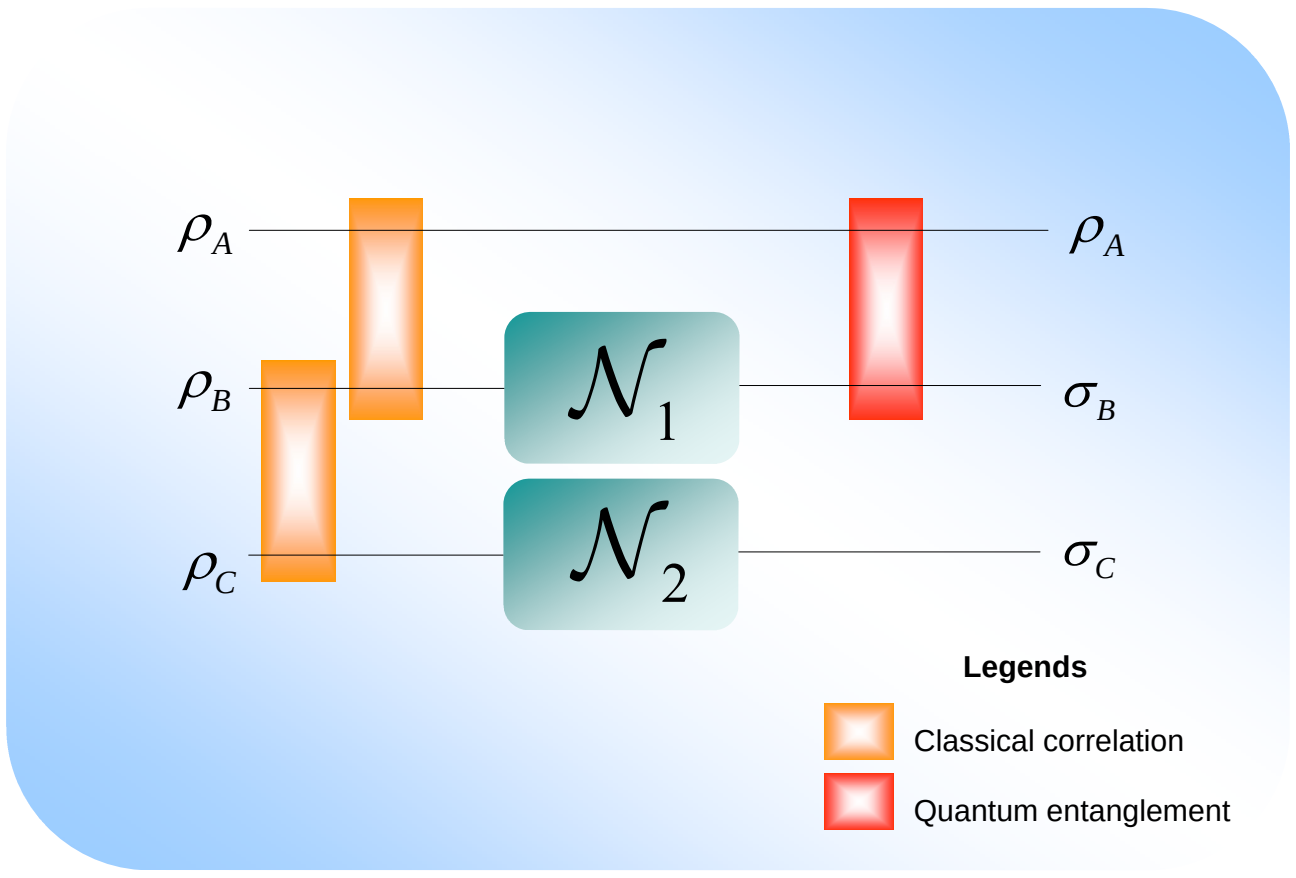

**Fig. 2.** The CC-property of quantum channels. Neither channel $\mathcal{N}_1$ nor $\mathcal{N}_2$ can transmit any quantum information or distillable entanglement (i.e., these channels are referred as classical-quantum channels); however, the noise transformation of the channels can generate distillable quantum entanglement between $\rho_A$ and $\sigma_B$ from classically correlated, unentangled inputs, $\rho_A$ and $\rho_B$.

To measure the amount of noise-generated distillable entanglement we consider the use of the $E(\cdot)$ *relative entropy of entanglement*, from the set of other entanglement measures [9,10], such as the negativity, concurrence or entanglement of formation [8,12]. By definition, the $E(\rho)$ relative entropy of entanglement function of the joint state $\rho$ of subsystems $A$ and $B$ is defined by the $D(\cdot\|\cdot)$ quantum relative entropy function, as

$$E(\rho) = \min_{\rho_{AB}} D(\rho\|\rho_{AB}) = \min_{\rho_{AB}} Tr(\rho\log\rho) - Tr(\rho\log(\rho_{AB})), \tag{2}$$

where $\rho_{AB}$ is the set of separable states $\rho_{AB} = \sum_{i=1}^{n} p_i \rho_{A,i} \otimes \rho_{B,i}$. As we have found, the following connection holds for the amount of noise-generated entanglement. *The achievable entanglement between* $\rho_A$ *and* $\sigma_B$ *is* $E(\sigma_{AB}) = \max_{v_+ - v_-}(v_+ - v_-)$, *where* $v_+, v_-$ *are the eigenvalues of* *channel output density matrix* $\sigma_{AB}$. We characterized an input system, for which the amount of entanglement between the *separable* input system $\rho_{AB}$ and the *entangled* channel *output* density matrix $\sigma_{AB}$ is

$$\begin{aligned} E\left(\sigma_{AB}\right) &= \min_{\rho_{AB}} D\left(\sigma_{AB} \| \rho_{AB}\right) \\ &= \left(\frac{1}{2} - \frac{1}{2}\left(v_{+} - v_{-}\right)\right) E\left(\left|\beta_{00}\right\rangle\left\langle\beta_{00}\right|\right) - \left(\frac{1}{2} - \frac{3}{2}\left(v_{+} - v_{-}\right)\right) E\left(\left|\beta_{10}\right\rangle\left\langle\beta_{10}\right|\right) \\ &= \left[\left(\frac{1}{2} - \frac{1}{2}\left(v_{+} - v_{-}\right)\right) - \left(\frac{1}{2} - \frac{3}{2}\left(v_{+} - v_{-}\right)\right)\right] E\left(\left|\Psi\right\rangle\left\langle\Psi\right|\right) \\ &= \max_{v_{+} - v_{-}}\left(v_{+} - v_{-}\right) \\ &= \left(1 - p\right) \cdot \left(v_{+} - v_{-}\right)_{in}, \end{aligned} \tag{3}$$

where $E\left(\left|\Psi\right\rangle\left\langle\Psi\right|\right) = E\left(\left|\beta_{00}\right\rangle\left\langle\beta_{00}\right|\right) = E\left(\left|\beta_{10}\right\rangle\left\langle\beta_{10}\right|\right) = 1$, $\left(v_{+} - v_{-}\right)_{in}$ is the difference of the eigenvalues in input system $\rho_{AB}$, $p$ is the noise of channel $\mathcal{N}_1$, while $\left|\beta_{00}\right\rangle = \frac{1}{\sqrt{2}}\left(\left|00\right\rangle + \left|11\right\rangle\right)$, $\left|\beta_{10}\right\rangle = \frac{1}{\sqrt{2}}\left(\left|00\right\rangle - \left|11\right\rangle\right)$ are the maximally entangled states. For the $E\left(\sigma_{AB}\right)$ relative entropy of entanglement of channel output system $\sigma_{AB}$ the inequality

$$0 < E\left(\sigma_{AB}\right) \leq \left(1 - p\right) \cdot \left(v_{+} - v_{-}\right)_{in} = \frac{2}{9} \tag{4}$$

holds, since $\left(1 - p\right) \leq \frac{2}{3}$ and $0 < \left(v_{+} - v_{-}\right)_{in} \leq \frac{1}{3}$. In Fig. 3, the amount of the noise-generated distillable entanglement $E\left(\sigma_{AB}\right)$ is summarized in the function of the difference of eigenvalues $v_{+}$ and $v_{-}$ of $\sigma_{AB}$.

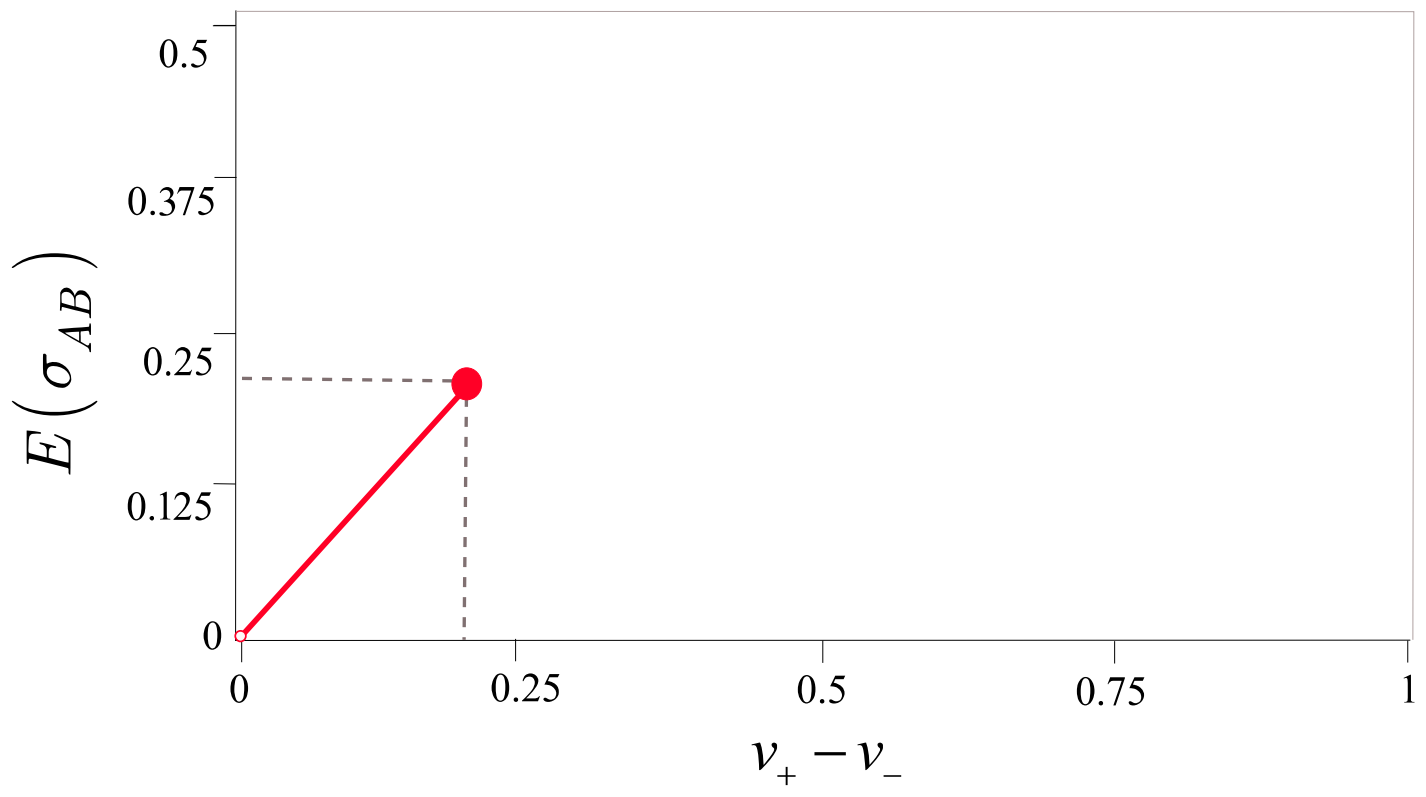

**Fig. 3.** The amount of noise-generated distillable entanglement in the function of the difference of the eigenvalues of the channel output density matrix.

Our results confirmed that the CC-property works for the most natural and simplest channel models—for example, the Pauli channels. We found a combination of two very simple channels, the so called phase flip channel $\mathcal{N}_1$ and the entanglement-breaking channel $\mathcal{N}_2$, that can transmit classical correlation only [14]. The $p \geq \frac{1}{3}$ error probability of the channel $\mathcal{N}_1$ results in $Q(\mathcal{N}_1) = 0$, the entanglement-breaking channel has also $Q(\mathcal{N}_2) = 0$ since it measures the input system and outputs a classically correlated density matrix [16]. However, they can be used to generate quantum entanglement. The result is *distillable* entanglement instead of bound (useless) entanglement, which would be a consequence of an entanglement-binding channel. Entanglement-binding channels could exist only in the set of qudit channels, and these channels generate bound-entangled (PPT – Positive Partial Transpose) output which allows no quantum communication, - however private classical communication is possible by PPT states. This fact, *per se*, immediately excludes the possibility of the existence of bound-entanglement in the proposed setting. For details and further derivation of the various correlation measures, see the *Supplementary Information*.

# Conclusion

In this work we first proved that distillable quantum entanglement can be produced by the noise transformation of classical (low fidelity) channels, the result of which has severe consequences for future quantum communications. Our results make it possible to generate distillable entanglement between distant points from classically correlated inputs over quantum channels that have no capability of transmitting quantum information. We developed a new idea which exploits the most natural property of the communication channels and which opens new dimensions in the fields of quantum communications. The solution does not require any non-local operation or local measurement by the parties, only the use of standard quantum channels. It also constrains us to revise our current knowledge about quantum channels and the nature of information itself. We have to reveal those deeply involved, currently hidden and uncharacterized possibilities that quantum information still holds.

## Acknowledgements

The author would like to thank Kamil Bradler and Tomasz Paterek for useful discussions and comments, and Yichen Huang for suggesting Ref. [16]. The results discussed above are supported by the grant TAMOP-4.2.1/B-09/1/KMR-2010-0002, 4.2.2.B-10/1--2010-0009 and COST Action MP1006.

## References

[1] W. J. Munro, K. A. Harrison, A. M. Stephens, S. J. Devitt, and K. Nemoto, Nature Photonics, 10.1038/nphoton.2010.213, (2010)

[2] L. Hanzo, H. Haas, S. Imre, D. O'Brien, M. Rupp, L. Gyongyosi, Wireless Myths, Realities, and Futures: From 3G/4G to Optical and Quantum Wireless, Proceedings of the IEEE, Volume: 100, Issue: *Special Centennial Issue*, pp. 1853-1888. (2012)

[3] S. Imre and L. Gyongyosi, *Advanced Quantum Communications - An Engineering Approach*, Publisher: Wiley-IEEE Press (New Jersey, USA), John Wiley & Sons, Inc., The Institute of Electrical and Electronics Engineers. Hardcover: 524 pages, ISBN-10: 1118002369, ISBN-13: 978-11180023, (2012).

[4] R.V. Meter, T. D. Ladd, W.J. Munro, K. Nemoto, System Design for a Long-Line Quantum Repeater, IEEE/ACM Transactions on Networking 17(3), 1002-1013, (2009).

[5] C. H. Bennett and G. Brassard. Quantum cryptography: Public key distribution and coin tossing. In Proc. IEEE International Conference on Computers, Systems, and Signal Processing, pages 175–179, (1984).

[6] A.K. Ekert. Quantum cryptography based on Bell's theorem. Physical Review Letters, 67(6):661–663, (1991).

[7] C. Elliott, D. Pearson, and G. Troxel. Quantum cryptography in practice. In Proc. SIGCOMM 2003. ACM, (2003).

[8] T. S. Cubitt, F. Verstraete, W. Dür, and J. I. Cirac, Phys. Rev. Lett. 91, 037902 (2003).

[9] T. K. Chuan, J. Maillard, K. Modi, T. Paterek, M. Paternostro, and M. Piani, Quantum discord bounds the amount of distributed entanglement, arXiv:1203.1268v3, Phys. Rev. Lett. 109, 070501 (2012).

[10] A. Streltsov, H. Kampermann, D. Bruß, Quantum cost for sending entanglement, arXiv:1203.1264v22012. Phys. Rev. Lett. 108, 250501 (2012)

[11] A. Kay, Resources for Entanglement Distribution via the Transmission of Separable States, arXiv:1204.0366v4, Phys. Rev. Lett. 109, 080503 (2012).

[12] J. Park, S. Lee, Separable states to distribute entanglement, arXiv:1012.5162v2, Int. J. Theor. Phys. 51 (2012) 1100-1110 (2010).

[13] N. J. Cerf, Quantum cloning and the capacity of the Pauli channel, arXiv:quant-ph/9803058v2, Phys.Rev.Lett. 84 4497 (2000).

[14] K. Bradler, P. Hayden, D. Touchette, M. M. Wilde, Trade-off capacities of the quantum Hadamard channels, Physical Review *A* 81, 062312 (2010).

[15] C. H. Bennett et al., PRA 54, 3824 (1996)

[16] Y. Huang, Quantum discord for two-qubit X states: Analytical formula with very small worst-case error, Phys. Rev. A 88, 014302 (2013).

[17] Laszlo Gyongyosi, Sandor Imre: Distillable Entanglement from Classical Correlation, Proceedings of SPIE Quantum Information and Computation XI, 2013.

# The Correlation Conversion Property of Quantum Channels

Laszlo Gyongyosi

[1]Quantum Technologies Laboratory, Department of Telecommunications
*Budapest University of Technology and Economics*
2 Magyar tudosok krt, Budapest, H-1117, Hungary
[2]Information Systems Research Group, Mathematics and Natural Sciences
*Hungarian Academy of Sciences*
Budapest, H-1518, Hungary

gyongyosi@hit.bme.hu

# Supplementary Information

## S.1 Theorems and Proofs

In the Supplementary Information we provide the theorems and proofs. First, we discuss properties of the channel structure, then we characterize the input system. Finally, we show the results on the channel output system.

### S.1.1 Channel System

First, we show that channels $\mathcal{N}_1$ and $\mathcal{N}_2$ can transmit classical correlation only.

**Proposition 1**. *The channels $\mathcal{N}_1$ and $\mathcal{N}_2$ in the joint structure $\mathcal{N}_1 \otimes \mathcal{N}_2$ can transmit only classical information.*

**First channel: the phase flip channel**

Channel $\mathcal{N}_1$ transmits Alice's input system $\rho_B$ and generates output system $\mathcal{N}_1(\rho_B) = \sigma_B$.

In the current work we demonstrate the results for a *phase flip* channel $\mathcal{N}_1$ with error probability $p = p_x + p_y + p_z$ where

$$p_x \geq \frac{1}{6}, p_y \geq \frac{1}{6}, p_z = 0, \text{ and } p \geq \frac{1}{3}, \tag{S.1}$$

characterize the noise transformation of the channel. For this parameterization, we get a channel that can transmit classical correlation only, since the channel has no quantum capacity [14]:

$$Q(\mathcal{N}_1) = 1 - 2\left(p_x + p_y + p_z + \sqrt{p_x}\sqrt{p_y} + \sqrt{p_x}\sqrt{p_z} + \sqrt{p_y}\sqrt{p_z}\right) = 0\,. \tag{S.2}$$

We use this channel as the first channel in the joint construction $\mathcal{N}_1 \otimes \mathcal{N}_2$. The noise parameters $p_x, p_y$ and $p_z = 0$ affect the eigenvalues of the input density matrix $\rho_{AB}$ as will be shown in Section 1.3.

**Second channel: the entanglement-breaking channel**

The second channel $\mathcal{N}_2$ in $\mathcal{N}_1 \otimes \mathcal{N}_2$ is the entanglement-breaking channel $\mathcal{N}_{EB}$. Giving an entangled system to input $A'$ of an entanglement-breaking channel $\mathcal{N}_{EB}$, it will destroy every entanglement on its output $B$. Formally, a noisy quantum channel $\mathcal{N}_{EB}$ is *entanglement-breaking* if for a half of a maximally entangled input $\left|\Psi\right\rangle_{AA'}$, the output of the channel is a separable state [37]. Let us assume that the maximally entangled input system of an $\mathcal{N}_{EB}$ entanglement-breaking channel is $\left|\Psi\right\rangle_{AA'} = \frac{1}{\sqrt{d}}\sum_{i=0}^{d-1}\left|i\right\rangle_A \left|i\right\rangle_{A'}$. The output of $\mathcal{N}_{EB}$ can be expressed as follows:

$$\mathcal{N}_{EB}\left(\left|\Psi\right\rangle\left\langle\Psi\right|_{AA'}\right) = \sum_x p_X(x)\rho_x^A \otimes \rho_x^B\,, \tag{S.3}$$

where $p_X(x)$ represents an arbitrary probability distribution, while $\rho_x^A$ and $\rho_x^B$ are the separable density matrices of the output system. The *noise-transformation* of an entanglement-breaking channel $\mathcal{N}_{EB}$ can be described as follows: *it performs a complete von Neumann measurement on its input system* $\rho$*, and outputs a classically correlated* (or an uncorrelated, depending on the measurement) *density matrix* $\sigma = \mathcal{N}_{EB}(\rho)$. It can be formalized as follows:

$$\mathcal{N}_{EB}(\rho) = \sum_x Tr\left\{\Pi_x \rho\right\}\sigma_x, \tag{S.4}$$

where $\left\{\Pi_x\right\}$ represents a POVM (Positive Operator Valued Measure) measurement on $\rho$ and $\sigma_x$ is the output density matrix of the channel [37]. Any entanglement-breaking channel $\mathcal{N}_{EB}$ can be decomposed into three parts: channel $\mathcal{N}_{EB}^1$ that acts as a noisy transformation on $\rho$, a measurement operator $\left\{\Pi_x\right\}$, and a second channel $\mathcal{N}_{EB}^2$, that outputs the density matrix $\sigma_x$:

$$\mathcal{N}_2 = \mathcal{N}_{EB}^1 \circ \Pi \circ \mathcal{N}_{EB}^2\,. \tag{S.5}$$

In our setting $\mathcal{N}_2 = \mathcal{N}_{EB}$, and the input of the channel is the flag $\rho_C$, from the classically correlated density matrix $\rho_{AC}$. After the $\mathcal{N}_2$ channel has got the flag $\rho_C$, measures it and outputs a density matrix $\sigma_C = \mathcal{N}_2(\rho_C)$,

$$\mathcal{N}_2\left(\rho_C\right) = \sum_x Tr\left\{\Lambda_C \rho_C\right\}\sigma_C, \quad \text{(S.6)}$$

where $\left\{\Lambda_C\right\}$ defines a projective measurement in the standard basis $\left\{|0\rangle, |1\rangle\right\}$, while the output flag system $\sigma_C$ is an arbitrary density matrix.

The decomposition of the entanglement-breaking channel $\mathcal{N}_2$ is depicted in Fig. S.1. It contains two $I$ ideal channels as $\mathcal{N}_{EB}^1$ and $\mathcal{N}_{EB}^2$, and a $\Lambda_C$ projective measurement, as follows its noisy evolution can be rewritten as

$$\mathcal{N}_2 = I \circ \Lambda_C \circ I\,. \quad \text{(S.7)}$$

The channel $\mathcal{N}_2$ measures the input flag system $\rho_C$, then outputs the density matrix $\sigma_C$. As the result of measurement flag system $C$, system $AB$ collapses into a well specified state. The output density matrix $\sigma_C$ contains the result of the measurement $\left\{\Lambda_C\right\}$, which will be referred as a *one-bit classical message* '0' or '1' that will inform Bob about the measurement result. Using the classical information from $\mathcal{N}_2$, Bob will be able to determine whether he received an entangled (distillable) or a classically correlated system $B$. The measurement $\left\{\Lambda_C\right\}$ of $\mathcal{N}_2$ and the identification processes together called *post-selection*. It is immediately follows that the classical information from $\mathcal{N}_2$ encoded in $\sigma_C$, is a required information to Bob to determine whether system $AB$ has become entangled, or not. If the post-selection process is successful then Bob localized entanglement to $AB$, and we will refer it as *entanglement-localization*.

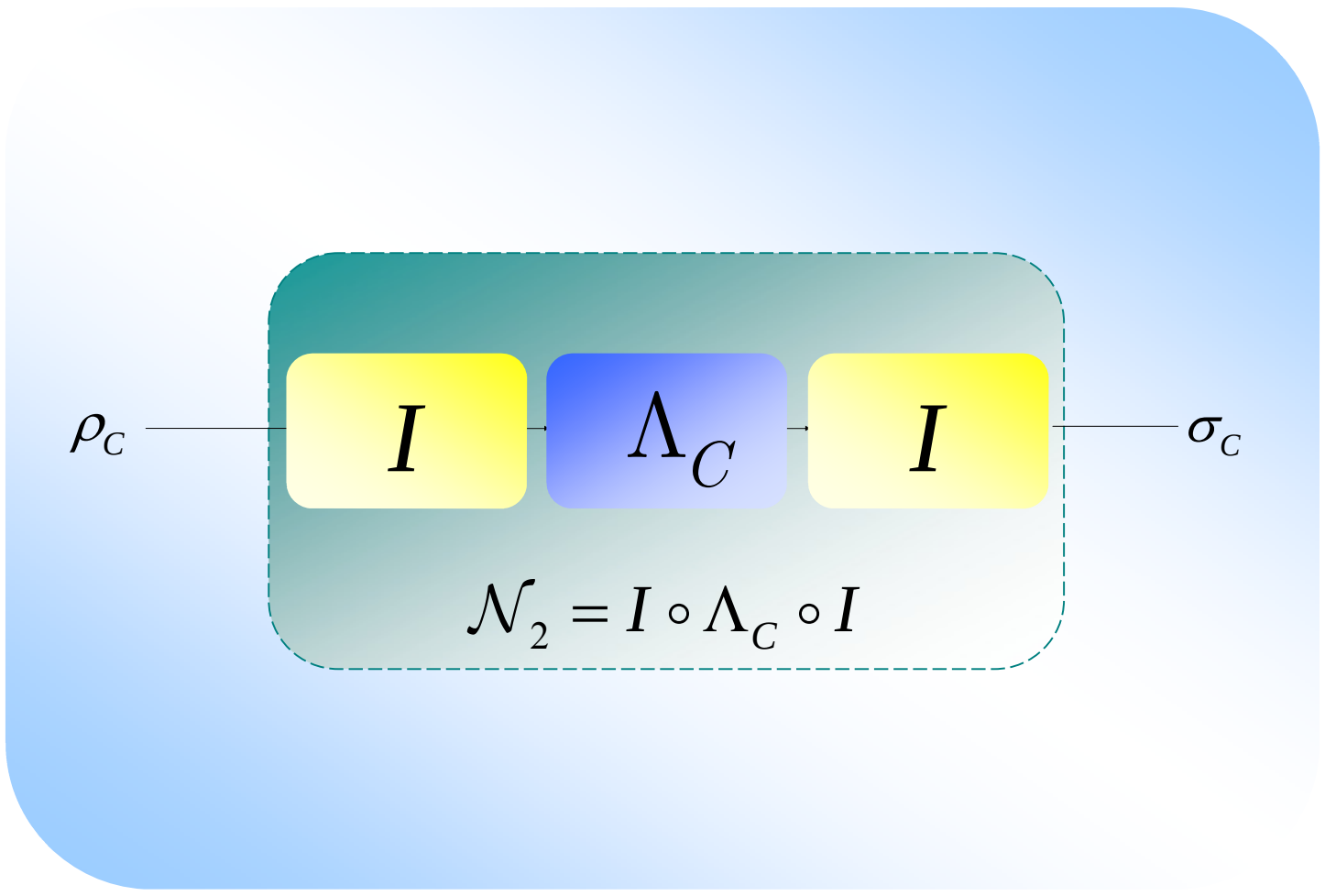

**Fig. S.1**. The decomposition of the entanglement-breaking channel $\mathcal{N}_2$. It measures the flag system $C$ and outputs the density matrix $\sigma_C$ to Bob, which encodes a classical bit (conditional state preparation). From the one-bit message, Bob will be able to identify the result of the $\Lambda_C$ projective measurement of the channel for the post-selection process.

The quantum capacity of any $\mathcal{N}_{EB}$ entanglement-breaking channels is trivially zero, since due to the $\{\Lambda_C\}$ measurement operator of the channel every entanglement vanishes. As follows, for $\mathcal{N}_2 = \mathcal{N}_{EB}$, after $\{\Lambda_C\}$ has been applied on $\rho_C$ we will have

$$Q(\mathcal{N}_2) = 0, \tag{S.8}$$

which makes no possible to transmit quantum entanglement over channel $\mathcal{N}_2$.

## Kraus Representation

The map of the quantum channel can also be expressed with a special representation called the *Kraus representation.* For a given input system $\rho_A$ and the quantum channel $\mathcal{N}$, this representation can be expressed as [4], [32-35]

$$\mathcal{N}(\rho_A) = \sum_i N_i \rho_A N_i^\dagger, \tag{S.9}$$

where $N_i$ are the Kraus operators, and $\sum_i N_i^\dagger N_i = I$. The isometric extension of $\mathcal{N}$ by means of the Kraus representation can be expressed as

$$\mathcal{N}(\rho_A) = \sum_i N_i \rho_A N_i^\dagger \to U_{A\to BE}(\rho_A) = \sum_i N_i \otimes |i\rangle_E. \tag{S.10}$$

The action of the quantum channel $\mathcal{N}$ on an operator $|k\rangle\langle l|$, where $\{|k\rangle\}$ is an orthonormal basis, also can be given in operator form using the Kraus operator $N_{kl} = \mathcal{N}(|k\rangle\langle l|)$. By exploiting the property $UU^\dagger = P_{BE}$, for the input quantum system $\rho_A$

$$U_{A\to BE}(\rho_A) = U\rho_A U^\dagger = \left(\sum_i N_i \otimes |i\rangle_E\right)\rho_A\left(\sum_j N_j^\dagger \otimes \langle j|_E\right) = \sum_{i,j} N_i \rho_A N_j^\dagger \otimes |i\rangle\langle j|_E. \tag{S.11}$$

Tracing out the environment, we get

$$Tr_E(U_{A\to BE}(\rho_A)) = \sum_i N_i \rho_A N_i^\dagger. \tag{S.12}$$

### Kraus Representation of the Phase Flip Channel

The effect of the phase flip channel $\mathcal{N}_1$ on the subsystem $\rho_B$ of $\rho_{AB}$ can be expressed in Kraus representation as follows [15-17], [41]:

$$\mathcal{N}_1(\rho_{AB}) = I(\rho_A) \otimes \mathcal{N}_1(\rho_B) = \sum_i N_i^{(B)} \rho_{AB} N_i^{(B)\dagger}, \tag{S.13}$$

where $I(\rho_A)$ denotes the identity transformation on subsystem $A$ and

$$\begin{aligned} N_0^{(B)} &= I_A \otimes diag\left(\sqrt{1-p/2}, \sqrt{1-p/2}\right), \\ N_1^{(B)} &= I_A \otimes diag\left(\sqrt{p/2}, -\sqrt{p/2}\right), \end{aligned} \tag{S.14}$$

while $p = p_x + p_y + p_z$ is the error probability of the channel $\mathcal{N}_1$.

**Kraus Representation of the Entanglement-breaking Channel**

The entanglement-breaking channel $\mathcal{N}_2$ on the subsystem $\rho_C$ of $\rho_{AC}$ can be expressed as

$$\mathcal{N}_2\left(\rho_{AC}\right) = I\left(\rho_A\right) \otimes \mathcal{N}_2\left(\rho_C\right) = \sum_i N_i^{(C)} \rho_{AC} N_i^{(C)\dagger}, \tag{S.15}$$

where

$$N_i^{(C)} = I_A \otimes \left|\xi_i\right\rangle_{C'} \left\langle \varsigma\right|_C, \tag{S.16}$$

where $C$ and $C'$ denote the input and output systems, and the Kraus-operators $N_i^{(C)}$ are unit rank. The sets $\left\{\left|\xi_i\right\rangle_{C'}\right\}$ and $\left\{\left|\varsigma\right\rangle_C\right\}$ each do not necessarily form an orthonormal set [37].

## S.1.2 Characterization of Input System

**Theorem 1**. *There exists input system $\rho_{ABC}$, that can be characterized by the $\left(v_+ - v_-\right)_{in}$ difference of the eigenvalues $v_+$, $v_-$ of the separable, classically correlated subsystem $\rho_{AB}$.*

*Proof.*

*Note*: *The results will be demonstrated for qubit channels.* Before the sending phase, Alice prepares the separable system $\rho_{ABC}$, which contain no quantum entanglement between $\rho_A$ and $\rho_B$. Alice holds $\rho_A$, while she feeds the systems $\rho_B$ and $\rho_C$, which will be the inputs of the joint channel structure $\mathcal{N}_1 \otimes \mathcal{N}_2$, where $\rho_B$ is the valuable system, and $\rho_C$ is a *flag* state.

The distillable entanglement will be prepared between systems $\rho_A$ and $\sigma_B = \mathcal{N}_1\left(\rho_B\right)$, after Bob has received the flag system $\sigma_C = \mathcal{N}_2\left(\rho_C\right)$. The process of decoherence on two qubit states has been exhaustively studied in the literature [15-31], [39], [41]. However, in our case the noise of the channel will affect only one system state, which still requires further investigation in the mathematical description.

The channel input system $\rho_{ABC}$ with the separable systems *A*, *B*, and flag state *C*, is prepared by Alice as follows[†]:

[†] *Note: The initial system in* (S.17) *contains no quantum entanglement between systems A and B, and will be referred as classically correlated. For the complete correctness, it is not pure classical correlation, since it has some positive quantum discord, see* (S.70). *We note that we are not interested in the further partitions (cuts)* [35-36] *of the initial state. To generate entanglement between A and B, local operations will be applied on B and C. These local operations make no possible to preserve entanglement in B and C, or in any partitions (cuts) of ABC that contain these subsystems.*

$$
\begin{aligned}
\rho_{ABC} = & \\
& \left(\frac{1}{4} - \frac{1}{4}\left(v_+ - v_-\right)_{in}\right)\left(|000\rangle\langle 000| + |000\rangle\langle 110| + |110\rangle\langle 000| + |110\rangle\langle 110|\right) + \\
& \left(\frac{1}{4} - \frac{3}{4}\left(v_+ - v_-\right)_{in}\right)\left(|000\rangle\langle 000| - |000\rangle\langle 110| - |110\rangle\langle 000| + |110\rangle\langle 110|\right) + \\
& \left(\frac{1}{2}\left(v_+ - v_-\right)_{in}\right)\left(|001\rangle\langle 001| + |011\rangle\langle 011| + |101\rangle\langle 101| + |111\rangle\langle 111|\right) \\
= & \\
& \left(\frac{1}{2} - \left(v_+ - v_-\right)_{in}\right)\left(|000\rangle\langle 000| + |110\rangle\langle 110|\right) + \\
& \left(\frac{1}{2}\left(v_+ - v_-\right)_{in}\right)\begin{pmatrix} |000\rangle\langle 110| + |110\rangle\langle 000| + |001\rangle\langle 001| + \\ |011\rangle\langle 011| + |101\rangle\langle 101| + |111\rangle\langle 111| \end{pmatrix},
\end{aligned}
\tag{S.17}
$$

where $\rho_{AB}$ is a separable Bell diagonal state [15-16], which can be expressed as

$$
\begin{aligned}
\rho_{AB} = & \left(\frac{1}{2} - \frac{1}{2}\left(v_+ - v_-\right)_{in}\right)\left(|00\rangle\langle 00| + |11\rangle\langle 11|\right) + \\
& \left(\frac{1}{2}\left(v_+ - v_-\right)_{in}\right)\left(|00\rangle\langle 11| + |11\rangle\langle 00|\right) + \\
& \left(\frac{1}{2}\left(v_+ - v_-\right)_{in}\right)\left(|01\rangle\langle 01| + |10\rangle\langle 10|\right).
\end{aligned}
\tag{S.18}
$$

where $v_+$, $v_-$ are the eigenvalues of density matrix $\rho_{AB}$ (will be defined in (S.24)) and $\left(v_+ - v_-\right)_{in} = \frac{1}{3}$ (the eigenvalues of the input system $\rho_{AB}$ are $v_+ = \frac{1}{2}$ and $v_- = \frac{1}{6}$), while the separable (from $\rho_{AB}$) mixed system $\rho_C$:

$$
\rho_C = \sum_i p_i |\psi_i\rangle\langle\psi_i| \tag{S.19}
$$

in the probabilistic mixture of the pure systems $|\psi_0\rangle = |0\rangle$ and $|\psi_1\rangle = |1\rangle$, is called the *flag.*

The noise of channel $\mathcal{N}_1$ will transform the eigenvalues into the range $0 < \left(v_+ - v_-\right) = \left(1 - p\right) \cdot \left(v_+ - v_-\right)_{in} \leq \frac{1}{3}$. To see that $AB$ and the flag $C$ together is a separable system, we also give here the density matrix of (S.17).

$$
\rho_{ABC} =
\begin{pmatrix}
\frac{1}{2}-\left(v_+-v_-\right)_{in} & 0 & 0 & 0 & 0 & 0 & \frac{1}{2}\left(v_+-v_-\right)_{in} & 0 \\
0 & \frac{1}{2}\left(v_+-v_-\right)_{in} & 0 & 0 & 0 & 0 & 0 & 0 \\
0 & 0 & 0 & 0 & 0 & 0 & 0 & 0 \\
0 & 0 & 0 & \frac{1}{2}\left(v_+-v_-\right)_{in} & 0 & 0 & 0 & 0 \\
0 & 0 & 0 & 0 & 0 & 0 & 0 & 0 \\
0 & 0 & 0 & 0 & 0 & \frac{1}{2}\left(v_+-v_-\right)_{in} & 0 & 0 \\
\frac{1}{2}\left(v_+-v_-\right)_{in} & 0 & 0 & 0 & 0 & 0 & \frac{1}{2}-\left(v_+-v_-\right)_{in} & 0 \\
0 & 0 & 0 & 0 & 0 & 0 & 0 & \frac{1}{2}\left(v_+-v_-\right)_{in}
\end{pmatrix},
\tag{S.20}
$$

where $\rho_{AB}$ was given in (S.18), and can be expressed in matrix form as:

$$\rho_{AB} = \begin{pmatrix} \frac{1}{2}-\frac{1}{2}\left(v_+-v_-\right)_{in} & 0 & 0 & \frac{1}{2}\left(v_+-v_-\right)_{in} \\ 0 & \frac{1}{2}\left(v_+-v_-\right)_{in} & 0 & 0 \\ 0 & 0 & \frac{1}{2}\left(v_+-v_-\right)_{in} & 0 \\ \frac{1}{2}\left(v_+-v_-\right)_{in} & 0 & 0 & \frac{1}{2}-\frac{1}{2}\left(v_+-v_-\right)_{in} \end{pmatrix}, \tag{S.21}$$

while $\left(v_+ - v_-\right)_{in}$ is the difference of the eigenvalues in input system $\rho_{AB}$. System $\rho_{AB}$ is clearly separable and contains no distillable entanglement, which can also be easily checked by the Peres-Horodecki criterion [31-32]: the partial transposes will be positive, i.e., $\left(\rho_{AB}\right)^{T_A} \geq 0$ and $\left(\rho_{AB}\right)^{T_B} \geq 0$, which trivially follows since $\rho_{AB}$ is a separable Bell diagonal state. The flag system $\rho_C$ is also separable and contains no quantum entanglement since the partial transpose of $\rho_{ABC}$ with respect to $C$ is positive, i.e., $\left(\rho_{ABC}\right)^{T_C} \geq 0$, see later in (S.34). The eigenvalues $v_+$, $v_-$ of matrix $\rho_{AB}$ can be expressed as follows. First, we rewrite system $\rho_{AB}$ in the following representation [15-22], [41]:

$$\rho_{AB} = \frac{1}{4}\left( I \otimes I + \mathbf{r} \cdot \vec{\sigma} \otimes I + I \otimes \mathbf{s} \cdot \vec{\sigma} + \sum_{i=1}^{3} c_i \sigma_i \otimes \sigma_i \right), \tag{S.22}$$

where $\mathbf{r}$ and $\mathbf{s}$ are the Bloch vectors, $\vec{\sigma} = \left[\sigma_x, \sigma_y, \sigma_z\right]$ with the Pauli matrices $\sigma_i$, while $c_i$ are real parameters $-1 \leq c_i \leq 1$. For a Bell diagonal state $\mathbf{r} = \mathbf{s} = 0$. Choosing $\mathbf{r} = \left(0, 0, r\right)$ and $\mathbf{s} = \left(0, 0, s\right)$, the input state in (S.22) can be given in matrix representation as follows:

$$\rho_{AB} = \frac{1}{4}\begin{pmatrix} 1+r+s+c_3 & 0 & 0 & c_1-c_2 \\ 0 & 1+r-s-c_3 & c_1+c_2 & 0 \\ 0 & c_1+c_2 & 1-r+s-c_3 & 0 \\ c_1-c_2 & 0 & 0 & 1-r-s+c_3 \end{pmatrix}. \tag{S.23}$$

Then, the eigenvalues $v_+$, $v_-$ of $\rho_{AB}$ are defined as

$$\begin{aligned} v_+ &= \frac{1}{4}\left(1 - c_3 + \sqrt{\left(r - s\right)^2 + \left(c_1 + c_2\right)^2}\right) \geq 0, \\ v_- &= \frac{1}{4}\left(1 - c_3 - \sqrt{\left(r - s\right)^2 + \left(c_1 + c_2\right)^2}\right) \geq 0. \end{aligned} \tag{S.24}$$

The other two eigenvalues $u_+$, $u_-$ can be defined as follows:

$$\begin{aligned} u_+ &= \frac{1}{4}\left(1 + c_3 + \sqrt{\left(r + s\right)^2 + \left(c_1 - c_2\right)^2}\right) \geq 0, \\ u_- &= \frac{1}{4}\left(1 + c_3 - \sqrt{\left(r + s\right)^2 + \left(c_1 - c_2\right)^2}\right) \geq 0. \end{aligned} \tag{S.25}$$

System $\rho_{AC}$ can be expressed in the same way, as

$$\rho_{AC} = \frac{1}{4}\begin{pmatrix} 1+r+s+c_3 & 0 & 0 & c_1-c_2 \\ 0 & 1+r-s-c_3 & c_1+c_2 & 0 \\ 0 & c_1+c_2 & 1-r+s-c_3 & 0 \\ c_1-c_2 & 0 & 0 & 1-r-s+c_3 \end{pmatrix}, \tag{S.26}$$

and the eigenvalues of this matrix will be denoted by

$$\begin{aligned} \kappa_+ &= \frac{1}{4}\left(1 - c_3 + \sqrt{\left(r-s\right)^2 + \left(c_1 + c_2\right)^2}\right) \geq 0, \\ \kappa_- &= \frac{1}{4}\left(1 - c_3 - \sqrt{\left(r-s\right)^2 + \left(c_1 + c_2\right)^2}\right) \geq 0, \end{aligned} \tag{S.27}$$

and

$$\begin{aligned} \tau_+ &= \frac{1}{4}\left(1 + c_3 + \sqrt{\left(r+s\right)^2 + \left(c_1 - c_2\right)^2}\right) \geq 0, \\ \tau_- &= \frac{1}{4}\left(1 + c_3 - \sqrt{\left(r+s\right)^2 + \left(c_1 - c_2\right)^2}\right) \geq 0, \end{aligned} \tag{S.28}$$

respectively. Using this representation form, the required conditions for the separability of the input system can be given as follows. For separable systems $AB$ and $AC$, the conditions

$$\max\left\{v_+, v_-, u_+, u_-\right\} \leq \frac{1}{2}, \tag{S.29}$$

and

$$\max\left\{\kappa_+, \kappa_-, \tau_+, \tau_-\right\} \leq \frac{1}{2}, \tag{S.30}$$

have to be satisfied. Furthermore, assuming a Bell diagonal state ( $r = 0, s = 0$ ), the condition

$$\left|c_1\right| + \left|c_2\right| + \left|c_3\right| \leq 1 \tag{S.31}$$

also trivially follows for the separability for each systems, $AB$ and $AC$.

■

**Corollary 1**. *The separability of input system* $\rho_{AB}$ for any $0 < \left(v_+ - v_-\right) \leq \frac{1}{3}$ is satisfied, since $\max\left\{v_+, v_-, u_+, u_-\right\} \leq \frac{1}{2}$.

**Remark 1.** (On the role of classical communication). *The proposed scheme uses only quantum channels between Alice and Bob and no classical channels applied in the process. The entanglement generation requires only the use of quantum channels and does not contain any non-local operation or classical communication between the parties. The post-selection process is also realized by itself the noise of quantum channel* $\mathcal{N}_2$. *The one-bit classical message is produced by the local measurement* $\left\{\Lambda_C\right\}$ *of* $\mathcal{N}_2$, *and the result will be communicated to Bob by* $\mathcal{N}_2$. *Alice does not send any classical information to Bob, nor Bob to Alice.*

*Note: The proposed scheme could be reduced to classical communication between Alice and Bob, if and only if in the input system* $\rho_{ABC}$ *the quantum discord would be* $\mathcal{D}\left(\rho_{ABC}\right) = 0$, *however*

*this not the case:* $\mathcal{D}\left(\rho_{ABC}\right) > 0$, *see later* (S.61), (S.65) and (S.70) *which makes no possible to interpret the transmission of C as classical communication* [9].

**Remark 2.** (On the impossibility of entanglement generation by LOCC). *We are interested in the entanglement between A and B. The theorem on the impossibility of entanglement generation by local operations* [38] *is not violated, because the local operations will be applied to B and C, instead of A and B. Channels* $\mathcal{N}_1$ *and* $\mathcal{N}_2$ *are CPTP (Completely Positive Trace Preserving) maps, which can be interpreted as local operations on systems B and C. The first channel* $\mathcal{N}_1$ *acts as a local operation on B, the entanglement-breaking channel* $\mathcal{N}_2$ *performs a local measurement* $\left\{\Lambda_C\right\}$ *on C, then conditionally prepares a density matrix depending on the measurement outcome (conditional state preparation). Since channel* $\mathcal{N}_2$ *sends the output density matrix only to Bob, channel* $\mathcal{N}_2$ *also represents a local operation.*

As the results have confirmed, distillable quantum entanglement can be generated only by the use of standard quantum channels $\mathcal{N}_1$ and $\mathcal{N}_2$, from which Corollary 2 follows.

**Corollary 2**. *Local operations on B and C can result in distillable quantum entanglement between A and B. These local operations are two CPTP maps, which makes no possible to preserve entanglement in subsystems B and C.*

## Required Conditions

*In input system* $\rho_{ABC}$ *the subsystem AB is classically correlated. The partial transposes of* $\rho_{AB}$ *with respect to the subsystems have to be positive.*

The input density matrix $\rho_{AB}$ has to be classically correlated and system $\rho_{ABC}$ has to be separable, which also can be given by different conditions. Using the Peres-Horodecki criterion [31-32] it is summarized as:

$$\begin{gathered}\left(\rho_{AB}\right)^{T_A} \geq 0,\\ \left(\rho_{AB}\right)^{T_B} \geq 0,\\ \left(\rho_{ABC}\right)^{T_B} \geq 0,\\ \left(\rho_{ABC}\right)^{T_C} \geq 0,\end{gathered} \tag{S.32}$$

hold true, and $0 < \left(v_+ - v_-\right) \leq \frac{1}{3}$ by the initial assumption on the input system.

**Proposition 2**. *These conditions on systems* $\rho_{ABC}$ *and* $\rho_{AB}$ *are satisfied in the initial state.*

These conditions will be checked by the Peres-Horodecki criterion [31-32], by taking the partial transposes $\left(\rho_{AB}\right)^{T_A}$, $\left(\rho_{AB}\right)^{T_B}$, $\left(\rho_{ABC}\right)^{T_B}$ and $\left(\rho_{ABC}\right)^{T_C}$ of the input system $\rho_{ABC}$ of (S.20). The positivity of $\left(\rho_{AB}\right)^{T_A}$ and $\left(\rho_{AB}\right)^{T_B}$ trivially follows from (S.18), since $\rho_{AB}$ is a separable Bell diagonal state. For simplicity we will show the partial transpose of $\rho_{ABC}$ with respect to *C*, where $\sigma_{ABC}$ (*before* $\left\{\Lambda_C\right\}$ *has applied to the flag* $\rho_C$) is:

$$\sigma_{ABC} = \begin{pmatrix} \frac{1}{2}-\left(v_+-v_-\right) & 0 & 0 & 0 & 0 & 0 & \frac{1}{2}\left(v_+-v_-\right) & 0 \\ 0 & \frac{1}{2}\left(v_+-v_-\right) & 0 & 0 & 0 & 0 & 0 & 0 \\ 0 & 0 & 0 & 0 & 0 & 0 & 0 & 0 \\ 0 & 0 & 0 & \frac{1}{2}\left(v_+-v_-\right) & 0 & 0 & 0 & 0 \\ 0 & 0 & 0 & 0 & 0 & 0 & 0 & 0 \\ 0 & 0 & 0 & 0 & 0 & \frac{1}{2}\left(v_+-v_-\right) & 0 & 0 \\ \frac{1}{2}\left(v_+-v_-\right) & 0 & 0 & 0 & 0 & 0 & \frac{1}{2}-\left(v_+-v_-\right) & 0 \\ 0 & 0 & 0 & 0 & 0 & 0 & 0 & \frac{1}{2}\left(v_+-v_-\right) \end{pmatrix}. \tag{S.33}$$

System $\left(\rho_{ABC}\right)^{T_C}$ can be expressed as follows:

$$\left(\rho_{ABC}\right)^{T_C} = \begin{pmatrix} \frac{1}{2}-\left(v_+-v_-\right)_{in} & 0 & 0 & 0 & 0 & 0 & 0 & 0 \\ 0 & \frac{1}{2}\left(v_+-v_-\right)_{in} & 0 & 0 & 0 & 0 & \frac{1}{2}\left(v_+-v_-\right)_{in} & 0 \\ 0 & 0 & 0 & 0 & 0 & 0 & 0 & 0 \\ 0 & 0 & 0 & \frac{1}{2}\left(v_+-v_-\right)_{in} & 0 & 0 & 0 & 0 \\ 0 & 0 & 0 & 0 & 0 & 0 & 0 & 0 \\ 0 & 0 & 0 & 0 & 0 & \frac{1}{2}\left(v_+-v_-\right)_{in} & 0 & 0 \\ 0 & \frac{1}{2}\left(v_+-v_-\right)_{in} & 0 & 0 & 0 & 0 & \frac{1}{2}-\left(v_+-v_-\right)_{in} & 0 \\ 0 & 0 & 0 & 0 & 0 & 0 & 0 & \frac{1}{2}\left(v_+-v_-\right)_{in} \end{pmatrix}, \tag{S.34}$$

where $v_+$, $v_-$ are the eigenvalues of the input density matrix $\rho_{AB}$. One readily can check by the Peres-Horodecki criterion [31-32] that the partial transpose is positive, hence

$$\left(\rho_{ABC}\right)^{T_C} \geq 0, \tag{S.35}$$

and

$$\left(\rho_{ABC}\right)^{T_B} \geq 0. \tag{S.36}$$

Tracing out flag system *C* from $\rho_{ABC}$, one can check easily that the partial transpose of the resulting matrix $Tr_C\left(\rho_{ABC}\right)$ with respect to *A* and *B* is positive, since $\left(\rho_{AB}\right)^{T_A} \geq 0$ and $\left(\rho_{AB}\right)^{T_B} \geq 0$. Since these conditions on $\rho_{ABC}$ are all satisfied, it also proves that in the separable input system *ABC*, system *AB* contains no quantum entanglement.

**Proposition 3.** *The noise of* $\mathcal{N}_1$ *affects the eigenvalues* $v_+, v_-$ *of* $\rho_{ABC}$. *The noise of* $\mathcal{N}_1$ *can transform the initial eigenvalues of* $\rho_{AB}$ *in the output system* $\sigma_{AB} = \rho_A \mathcal{N}_1(\rho_B)$, *as such* $0 < (v_+ - v_-) \leq \frac{2}{9}$ *will hold. In this domain positive quantum entanglement can be generated between* $\rho_A$ *and channel output* $\mathcal{N}_1(\rho_B) = \sigma_B$.

# S.1.3 The Correlation Conversion Property

*In the output system* $\sigma_{ABC} = \rho_A \mathcal{N}_1(\rho_B) \otimes \mathcal{N}_2(\rho_C)$ *of* $\mathcal{N}_1 \otimes \mathcal{N}_2$ *two conditions have to be satisfied. First, the flag system C has to be separable from systems A and B. Second, for positive quantum entanglement in* $\sigma_{AB}$ *the difference between the eigenvalues* $v_+, v_-$ *of output matrix* $\sigma_{AB}$, *the condition* $0 < (v_+ - v_-)$ *has to hold.*

The *Correlation Conversion property* of quantum channels is summarized in Theorem 2.

**Theorem 2**. (On the Correlation Conversion property of quantum channels). *There exists channels* $\mathcal{N}_1$ *and* $\mathcal{N}_2$ *which can generate distillable entanglement from classically correlated input* $\rho_{AB}$, *between systems* $\rho_A$ *and channel output* $\sigma_B = \mathcal{N}_1(\rho_B)$, *where neither channel* $\mathcal{N}_1$, *nor* $\mathcal{N}_2$ *can transmit any quantum entanglement,* $Q(\mathcal{N}_1) = Q(\mathcal{N}_2) = 0$. *The noise transformation of the channel can retransform the density matrix in such a way that it results in distillable entanglement between systems A and B.*

*Proof.*
Here we prove that the output system of $\mathcal{N}_1 \otimes \mathcal{N}_2$ contains quantum entanglement between Alice's density matrix $\rho_A$ and the channel output $\sigma_B$. According to the Theorem 2, the noise of channel system $\mathcal{N}_1 \otimes \mathcal{N}_2$ generates quantum entanglement between Alice's density matrix $\rho_A$ and channel output $\sigma_B$ from the classically correlated input systems $\rho_{AB}$ and $\rho_{AC}$. After Bob has received systems $\sigma_B$ and $\sigma_C$, the resulting system state will be referred as follows:

$$\sigma_{ABC} = \rho_A \mathcal{N}_1(\rho_B) \otimes \mathcal{N}_2(\rho_C), \tag{S.37}$$

in which system the flag $C$ remains separable, since the partial transposes of $\sigma_{ABC}$ are non-negative, see (S.34), and the $v_+$, $v_-$ eigenvalues of density matrix $\sigma_{AB}$ affected by the noise of $\mathcal{N}_1 \otimes \mathcal{N}_2$, and $(v_+ - v_-) = (1-p) \cdot (v_+ - v_-)_{in}$ with relations

$$0 < (v_+ - v_-) \leq \frac{2}{9}, \tag{S.38}$$

and

$$1 - 2(v_+ + v_-) + 2(v_+ - v_-) = 1. \tag{S.39}$$

After the flag system $C$ has been removed (since it was fed to the entanglement-breaking channel $\mathcal{N}_2$), the system state reduces to

$$Tr_C\left(\sigma_{ABC}\right) = \sigma_{AB}. \tag{S.40}$$

The density matrix between Alice's system $\rho_A$ and channel output $\sigma_B$ can be expressed as follows (*before channel* $\mathcal{N}_2$ *has applied* $\left\{\Lambda_C\right\}$ *to the flag* $\rho_C$):

$$\sigma_{AB} = \begin{pmatrix} \frac{1}{2}-\frac{1}{2}\left(v_+-v_-\right) & 0 & 0 & \frac{1}{2}\left(v_+-v_-\right) \\ 0 & \frac{1}{2}\left(v_+-v_-\right) & 0 & 0 \\ 0 & 0 & \frac{1}{2}\left(v_+-v_-\right) & 0 \\ \frac{1}{2}\left(v_+-v_-\right) & 0 & 0 & \frac{1}{2}-\frac{1}{2}\left(v_+-v_-\right) \end{pmatrix}, \tag{S.41}$$

where $v_+$, $v_-$ are the eigenvalues of the channel output density matrix $\sigma_{AB}$, and $0 < \left(v_+ - v_-\right) \leq \frac{2}{9}$, according to the characterization of the input system $\rho_{AB}$. One can further readily check that matrix $\sigma_{AB}$ in (S.41) has no negative partial transpose, which shows that $\rho_A$ and $\sigma_B$ still have not become entangled: $\left(\rho_{AB}\right)^{T_A}$, $\left(\rho_{AB}\right)^{T_B} \geq 0$. To achieve the entanglement in $AB$, the matrix (S.41) has to be decomposable into two different matrices, and its decomposition in determined by the flag system $C$. This *post-selection* process [8-12], [36-37] will be made by the entanglement-breaking channel $\mathcal{N}_2$. It will be possible if and only if the flag system $C$ has been transmitted over $\mathcal{N}_2$, and after $B$ has been received by Bob, i.e., there is a *causality* in the post-selection process: the flag $C$ cannot be measured by $\mathcal{N}_2$ before Bob would have not received $B$ from $\mathcal{N}_1$. On the other hand, without any information from $\mathcal{N}_2$, Bob will not be able to determine whether he received an entangled system $B$, or he owns just a classically correlated system. *The entanglement-breaking channel* $\mathcal{N}_2$ *will give the answer to Bob.* The output of $\mathcal{N}_2$ is a one-bit classical message that informs Bob about the result [42].

The flag system $\rho_C$ will be fed to the input of the entanglement-breaking channel $\mathcal{N}_2$, with $Q\left(\mathcal{N}_2\right) = 0$ and $C\left(\mathcal{N}_2\right) > 0$. The input flag $\rho_C$ is assumed to be in the probabilistic mixture of the pure systems $\left|\psi_0\right\rangle = \left|0\right\rangle$ and $\left|\psi_1\right\rangle = \left|1\right\rangle$, hence the output of will $C$=0 or $C$=1, after the channel $\mathcal{N}_2$ has applied the measurement operator $\left\{\Lambda_C\right\}$ to $\rho_C$, using the standard basis $\left\{\left|0\right\rangle,\left|1\right\rangle\right\}$. Channel $\mathcal{N}_2$ can be decomposed as $\mathcal{N}_{EB}^1 = I$ and $\mathcal{N}_{EB}^2 = I$, and a $\left\{\Lambda_C\right\}$ projective measurement $\mathcal{N}_2 = I \circ \Lambda_C \circ I$. After the flag system $\rho_C$ has been transmitted over $\mathcal{N}_2$, it will simply be traced out by the partial transpose operator $Tr_C\left(\cdot\right)$ and the final system state will reduce to $Tr_C\left(\sigma_{ABC}\right) = \sigma_{AB}$. The flag state has no impact on the amount of the generated entanglement over $\mathcal{N}_1$ in $\sigma_{AB}$. On the other hand, the measurement $\left\{\Lambda_C\right\}$ of $\mathcal{N}_2$ is a probabilistic process, which causes a decrease in the amount of generable entanglement, as will be quantified in *Theorem 3.*

**Remark 3**. *The output of the $\mathcal{N}_2$ is a necessary condition to achieve entanglement in $\sigma_{AB}$. Before the output of the entanglement-breaking channel $\mathcal{N}_2$ the localization of entanglement is not possible since the $\sigma_{AB}$ matrix is in the in the probabilistic mixture of the two possible systems $\sigma_{AB} = (\sigma_{AB})_0 + (\sigma_{AB})_1$, where 0 and 1 is the one-bit classical output of channel $\mathcal{N}_2$.*

After the channel $\mathcal{N}_2$ has applied $\{\Lambda_C\}$ to the flag system *C*, the output system $\sigma_{ABC}$ in (S.33) can be rewritten as follows:

$$\sigma_{ABC} = (\sigma_{AB})_0 \otimes |0\rangle\langle 0|_C + (\sigma_{AB})_1 \otimes |1\rangle\langle 1|_C, \tag{S.42}$$

and can be decomposed as:

$$\sigma_{ABC} =$$

$$\begin{pmatrix} \frac{1}{2}-(v_+-v_-) & 0 & 0 & 0 & 0 & 0 & \frac{1}{2}(v_+-v_-) & 0 \\ 0 & 0 & 0 & 0 & 0 & 0 & 0 & 0 \\ 0 & 0 & 0 & 0 & 0 & 0 & 0 & 0 \\ 0 & 0 & 0 & 0 & 0 & 0 & 0 & 0 \\ 0 & 0 & 0 & 0 & 0 & 0 & 0 & 0 \\ 0 & 0 & 0 & 0 & 0 & 0 & 0 & 0 \\ \frac{1}{2}(v_+-v_-) & 0 & 0 & 0 & 0 & 0 & \frac{1}{2}-(v_+-v_-) & 0 \\ 0 & 0 & 0 & 0 & 0 & 0 & 0 & 0 \end{pmatrix} + \begin{pmatrix} 0 & 0 & 0 & 0 & 0 & 0 & 0 & 0 \\ 0 & \frac{1}{2}(v_+-v_-) & 0 & 0 & 0 & 0 & 0 & 0 \\ 0 & 0 & 0 & 0 & 0 & 0 & 0 & 0 \\ 0 & 0 & 0 & \frac{1}{2}(v_+-v_-) & 0 & 0 & 0 & 0 \\ 0 & 0 & 0 & 0 & 0 & 0 & 0 & 0 \\ 0 & 0 & 0 & 0 & 0 & \frac{1}{2}(v_+-v_-) & 0 & 0 \\ 0 & 0 & 0 & 0 & 0 & 0 & 0 & 0 \\ 0 & 0 & 0 & 0 & 0 & 0 & 0 & \frac{1}{2}(v_+-v_-) \end{pmatrix}. \tag{S.43}$$

From it follows that system $\sigma_{AB}$ in (S.41) can be decomposed into

$$\sigma_{AB} = (\sigma_{AB})_0 + (\sigma_{AB})_1 =$$

$$\begin{pmatrix} \frac{1}{2}-(v_+-v_-)+\frac{1}{2}(v_+-v_-) & 0 & 0 & \frac{1}{2}(v_+-v_-) \\ 0 & \frac{1}{2}(v_+-v_-) & 0 & 0 \\ 0 & 0 & \frac{1}{2}(v_+-v_-) & 0 \\ \frac{1}{2}(v_+-v_-) & 0 & 0 & \frac{1}{2}-(v_+-v_-)+\frac{1}{2}(v_+-v_-) \end{pmatrix}, \tag{S.44}$$

where

$$(\sigma_{AB})_0 = \begin{pmatrix} \frac{1}{2}-(v_+-v_-) & 0 & 0 & \frac{1}{2}(v_+-v_-) \\ 0 & 0 & 0 & 0 \\ 0 & 0 & 0 & 0 \\ \frac{1}{2}(v_+-v_-) & 0 & 0 & \frac{1}{2}-(v_+-v_-) \end{pmatrix} \tag{S.45}$$

and

$$(\sigma_{AB})_1 = \begin{pmatrix} \frac{1}{2}(v_+-v_-) & 0 & 0 & 0 \\ 0 & \frac{1}{2}(v_+-v_-) & 0 & 0 \\ 0 & 0 & \frac{1}{2}(v_+-v_-) & 0 \\ 0 & 0 & 0 & \frac{1}{2}(v_+-v_-) \end{pmatrix}. \tag{S.46}$$

Due to the measurement $\{\Lambda_C\}$ on *C* of the channel $\mathcal{N}_2$, system $\sigma_{AB}$ in (S.42) collapses into (S.45) or (S.46). If $\mathcal{N}_2$ measured *C*=0, then the entanglement-localization was successful, and Bob in the *post-selection* process will be able to use the entangled system *B*, after he received the output $\sigma_C$ from $\mathcal{N}_2$. During the process the flag system $C$ is trivially separable in $\sigma_{ABC}$ from the remaining parts, $\sigma_{AB}$. Moreover, the partial transposes $(\sigma_{AB})^{T_A}$, $(\sigma_{AB})^{T_B}$, $(\sigma_{ABC})^{T_B}$, $(\sigma_{ABC})^{T_C}$ are both still non-negative. On the other hand, after $\{\Lambda_C\}$ has been applied on $C$ by $\mathcal{N}_2$, the partial transposes of $(\sigma_{AB})_0$ will be *negative*: $((\sigma_{AB})_0)^{T_A} < 0$, $((\sigma_{AB})_0)^{T_B} < 0$, which makes possible to achieve entanglement between $A$ and $B$. The systems $(\sigma_{AB})_0$ or $(\sigma_{AB})_1$ cannot be post-selected without the output of the entanglement-breaking channel $\mathcal{N}_2$. The selection of system $(\sigma_{AB})_0$ in $\sigma_{AB}$, i.e., the *localization of entanglement* into $AB$ could not be made until the output of the entanglement-breaking channel $\mathcal{N}_2$ has not received by Bob, only their probabilistic mixture $\sigma_{AB} = (\sigma_{AB})_0 + (\sigma_{AB})_1$ exists for Bob. After the channel $\mathcal{N}_2$ has applied $\{\Lambda_C\}$ on the flag $C$, the entangled system $(\sigma_{AB})_0$ can be post-selected by Bob, pending the classical information from $\sigma_C$.

*Note:* In the input system $\rho_{AB}$ the *density matrix* $(\rho_{AB})_0$ could be selected by Alice if and only if she would have applied a measurement operator $\{\Lambda_C\}$ on $C$. However at that initial stage the flag $C$ cannot be measured, she can send only the classically correlated system $\rho_{AB} = (\rho_{AB})_0 + (\rho_{AB})_1$ to Bob. Assuming the case that Alice would apply a measurement $\{\Lambda_C\}$ on the flag $C$ in the initial phase (before the transmission) to get the entangled density matrix $(\rho_{AB})_0$, she will find that she is not able to send the entangled $B$ to Bob over $\mathcal{N}_1$, since $Q(\mathcal{N}_1) = 0$. It is also not possible over $\mathcal{N}_2$, because $Q(\mathcal{N}_2) = 0$ by the initial assumptions on $\mathcal{N}_1$ and $\mathcal{N}_2$. As follows, in the input system $\rho_{AB}$, only the partial transpose of $\rho_{AB}$ can be used to analyze the entanglement in $AB$, which is positive.

■

**Corollary 3**. *The partial transposes* $((\sigma_{AB})_0)^{T_A} < 0$, $((\sigma_{AB})_0)^{T_B} < 0$ *are negative in the channel output system* $\sigma_{AB}$.

While for the input system $(\rho_{AB})^{T_A} \geq 0$ and $(\rho_{AB})^{T_B} \geq 0$, see (S.21), in (S.42) the partial transposes $((\sigma_{AB})_0)^{T_A}$, $((\sigma_{AB})_0)^{T_B}$ of $(\sigma_{AB})_0$ are negative, see (S.45).

**Remark 4**. *Entanglement generation over* $\mathcal{N}_1$ *is possible if and only if the output of the entanglement-breaking channel* $\mathcal{N}_2$ *has been received by Bob. With no output from* $\mathcal{N}_2$, *the channel output AB would be* (S.41) *which system state would not result in distillable entanglement be-* *tween A and B. If Bob receives 0 from $\mathcal{N}_2$, then he will know that he received the entangled system $(\sigma_{AB})_0$, however the measurement of the entanglement-breaking channel $\mathcal{N}_2$ is a probabilistic process; $\mathcal{N}_2$ will decrease the amount of maximally generated entanglement over $\mathcal{N}_1$, as will be exactly quantified by the relative entropy of entanglement function in Theorem 3.*

The proposed channel output system $\sigma_{ABC}$ satisfies the separability requirements and the condition for the entanglement of $\rho_A$ and $\sigma_B$. As follows, the noise of channel structure $\mathcal{N}_1 \otimes \mathcal{N}_2$ can transform the input density matrices $\rho_B$ and $\rho_C$ in such a way that results in quantum entanglement between Alice's system $\rho_A$ and channel output $\sigma_B = \mathcal{N}_1(\rho_B)$.

**Corollary 4**. *The noise of channel $\mathcal{N}_1$ can transform the eigenvalues $v_+, v_-$ of $\rho_{AB}$ in such a way that $0 < (v_+ - v_-) \leq \frac{2}{9}$ in the channel output system $\sigma_{AB}$ is satisfied, and AB becomes entangled.*

The channel output system $\sigma_{AB}$ can also be expressed as follows:

$$\sigma_{AB} = \frac{1}{4}\left(I \otimes I + \mathbf{r} \cdot \sigma_z \otimes I + I \otimes \mathbf{s} \cdot \sigma_z + (1-p)c_1\sigma_x \otimes \sigma_x + (1-p)c_2\sigma_y \otimes \sigma_y + c_3\sigma_z \otimes \sigma_z\right), \tag{S.47}$$

which can be expressed in matrix representation as [15]:

$$\sigma_{AB} = \frac{1}{4}\begin{pmatrix} 1+r+s+c_3 & 0 & 0 & (1-p)c_1 - (1-p)c_2 \\ 0 & 1+r-s-c_3 & (1-p)c_1 + (1-p)c_2 & 0 \\ 0 & (1-p)c_1 + (1-p)c_2 & 1-r+s-c_3 & 0 \\ (1-p)c_1 - (1-p)c_2 & 0 & 0 & 1-r-s+c_3 \end{pmatrix}, \tag{S.48}$$

where $p$ is the error probability $p = p_x + p_y + p_z$ of channel $\mathcal{N}_1$. Due to the noise of $\mathcal{N}_1$, the eigenvalues $v_+$, $v_-$ of $\sigma_{AB}$ are changed from the initial values to

$$\begin{aligned} v_+ &= \frac{1}{4}\left(1 - c_3 + \sqrt{(r-s)^2 + \left((1-p)c_1 + (1-p)c_2\right)^2}\right), \\ v_- &= \frac{1}{4}\left(1 - c_3 - \sqrt{(r-s)^2 + \left((1-p)c_1 + (1-p)c_2\right)^2}\right), \end{aligned} \tag{S.49}$$

satisfying the required condition

$$0 < (v_+ - v_-) \leq \frac{2}{9}. \tag{S.50}$$

The other two eigenvalues $u_+$, $u_+$ of $\sigma_{AB}$ are irrelevant in the further calculations, since they have no effect on the amount of noise-generated entanglement.

## Capacity Calculations

Next we discuss the amount of distillable quantum entanglement in $\sigma_{AB}$ which can be produced by the noise of $\mathcal{N}_1 \otimes \mathcal{N}_2$, assuming the previously-shown input system characterization.

**Theorem 3**. (On the amount of noise-generated entanglement). *The relative entropy of entanglement between the classically correlated input system $\rho_{AB}$ and the output system $\sigma_{AB}$ is*

$$E\left(\sigma_{AB}\right) = \min_{\rho_{AB}} D\left(\sigma_{AB} \| \rho_{AB}\right) = \max_{v_+ - v_-}\left(v_+ - v_-\right) = \left(1-p\right)\cdot\left(v_+ - v_-\right)_{in},$$

*where $E\left(\sigma_{AB}\right)$ is the relative entropy of entanglement, $D\left(\cdot\|\cdot\right)$ is the relative entropy function, $\left(v_+ - v_-\right)_{in}$ is the difference of eigenvalues in $\rho_{AB}$, $p$ is the noise of the channel $\mathcal{N}_1$, while $v_+, v_-$ are the eigenvalues of channel output density matrix $\sigma_{AB}$.*

*Proof.*

First we show that the entanglement generated by $\mathcal{N}_1 \otimes \mathcal{N}_2$ can be measured by the quantum relative entropy function $D\left(\cdot\|\cdot\right)$. Then we prove that the amount of achievable quantum entanglement is determined by the noise characteristic of $\mathcal{N}_1 \otimes \mathcal{N}_2$. To measure the amount of entanglement we consider using the $E\left(\cdot\right)$ *relative entropy of entanglement* function [10-11], from the set of other entanglement measures, such as the negativity, concurrence or entanglement of formation [9, 13]. By definition, the $E\left(\rho\right)$ relative entropy of entanglement function of the joint state $\rho$ of subsystems $A$ and $B$ is defined by the quantum relative entropy function $D\left(\rho\|\rho_{AB}\right) = Tr\left(\rho\log\rho\right) - Tr\left(\rho\log\left(\rho_{AB}\right)\right)$, as

$$E\left(\rho\right) = \min_{\rho_{AB}} D\left(\rho\|\rho_{AB}\right), \tag{S.51}$$

where $\rho_{AB}$ the set of separable states $\rho_{AB} = \sum_{i=1}^{n} p_i \rho_{A,i} \otimes \rho_{B,i}$. *The amount of the noise-generated entanglement between $\rho_A$ and $\sigma_B$ is expressed by $E\left(\sigma_{AB}\right) = \max_{v_+ - v_-}\left(v_+ - v_-\right)$, where* $0 < E\left(\sigma_{AB}\right) \leq \frac{2}{9}$. The $E\left(\sigma_{AB}\right)$ relative entropy of entanglement between the separable channel input $\rho_{AB}$ and the channel output density matrix $\sigma_{AB}$ is

$$\begin{aligned} E\left(\sigma_{AB}\right) &= \min_{\rho_{AB}} D\left(\sigma_{AB}\|\rho_{AB}\right) \\ &= \left(\frac{1}{2} - \frac{1}{2}\left(v_+ - v_-\right)\right) E\left(\left|\beta_{00}\right\rangle\left\langle\beta_{00}\right|\right) - \left(\frac{1}{2} - \frac{3}{2}\left(v_+ - v_-\right)\right) E\left(\left|\beta_{01}\right\rangle\left\langle\beta_{01}\right|\right) \\ &= \left[\left(\frac{1}{2} - \frac{1}{2}\left(v_+ - v_-\right)\right) - \left(\frac{1}{2} - \frac{3}{2}\left(v_+ - v_-\right)\right)\right] E\left(\left|\Psi\right\rangle\left\langle\Psi\right|\right) \\ &= \left(v_+ - v_-\right)\cdot E\left(\left|\Psi\right\rangle\left\langle\Psi\right|\right) \\ &= \max_{v_+ - v_-}\left(v_+ - v_-\right) \\ &= \left(1-p\right)\cdot\left(v_+ - v_-\right)_{in}, \end{aligned} \tag{S.52}$$

where $\left(v_+ - v_-\right)_{in}$ is the difference of the eigenvalues in input system $\rho_{AB}$, and

$$E\left(\left|\Psi\right\rangle\left\langle\Psi\right|\right) = E\left(\left|\beta_{00}\right\rangle\left\langle\beta_{00}\right|\right) = E\left(\left|\beta_{01}\right\rangle\left\langle\beta_{01}\right|\right) = 1, \tag{S.53}$$

while $\left|\beta_{00}\right\rangle = \frac{1}{\sqrt{2}}\left(\left|00\right\rangle + \left|11\right\rangle\right)$ and $\left|\beta_{10}\right\rangle = \frac{1}{\sqrt{2}}\left(\left|00\right\rangle - \left|11\right\rangle\right)$ are the maximally entangled states. From the results on the $E\left(\sigma_{AB}\right)$ relative entropy of entanglement in the output system $\sigma_{AB}$, the inequality

$$0 < E\left(\sigma_{AB}\right) \leq \left(1-p\right)\cdot\left(v_+ - v_-\right)_{in} = \frac{2}{9} \tag{S.54}$$

trivially follows, since $\left(1-p\right) \leq \frac{2}{3}$ and $0 < \left(v_+ - v_-\right)_{in} \leq \frac{1}{3}$. In the separable input system $\rho_{AB} = \sum_{i=1}^{n} p_i \rho_{A,i} \otimes \rho_{B,i}$, there is no entanglement; the relative entropy of entanglement of $\rho_{AB}$ is

$$E\left(\rho_{AB}\right) = \min_{\rho_{AB}} D\left(\rho_{AB} \middle\| \sum_{i=1}^{n} p_i \rho_{A,i} \otimes \rho_{B,i}\right) = 0. \tag{S.55}$$

According to the Theorem 3, $\min_{\rho_{AB}} D\left(\sigma_{AB} \middle\| \rho_{AB}\right)$ taken between $\rho_{AB}$ and $\sigma_{AB}$ is analogous to the maximized difference $\max_{v_+ - v_-}\left(v_+ - v_-\right)$ of the eigenvalues $v_+, v_-$ of output matrix $\sigma_{AB}$.

In Fig. S.2, the $E\left(\sigma_{AB}\right)$ amount of noise-generated entanglement is summarized in function of the difference of eigenvalues $v_+$ and $v_-$ of $\sigma_{AB}$.

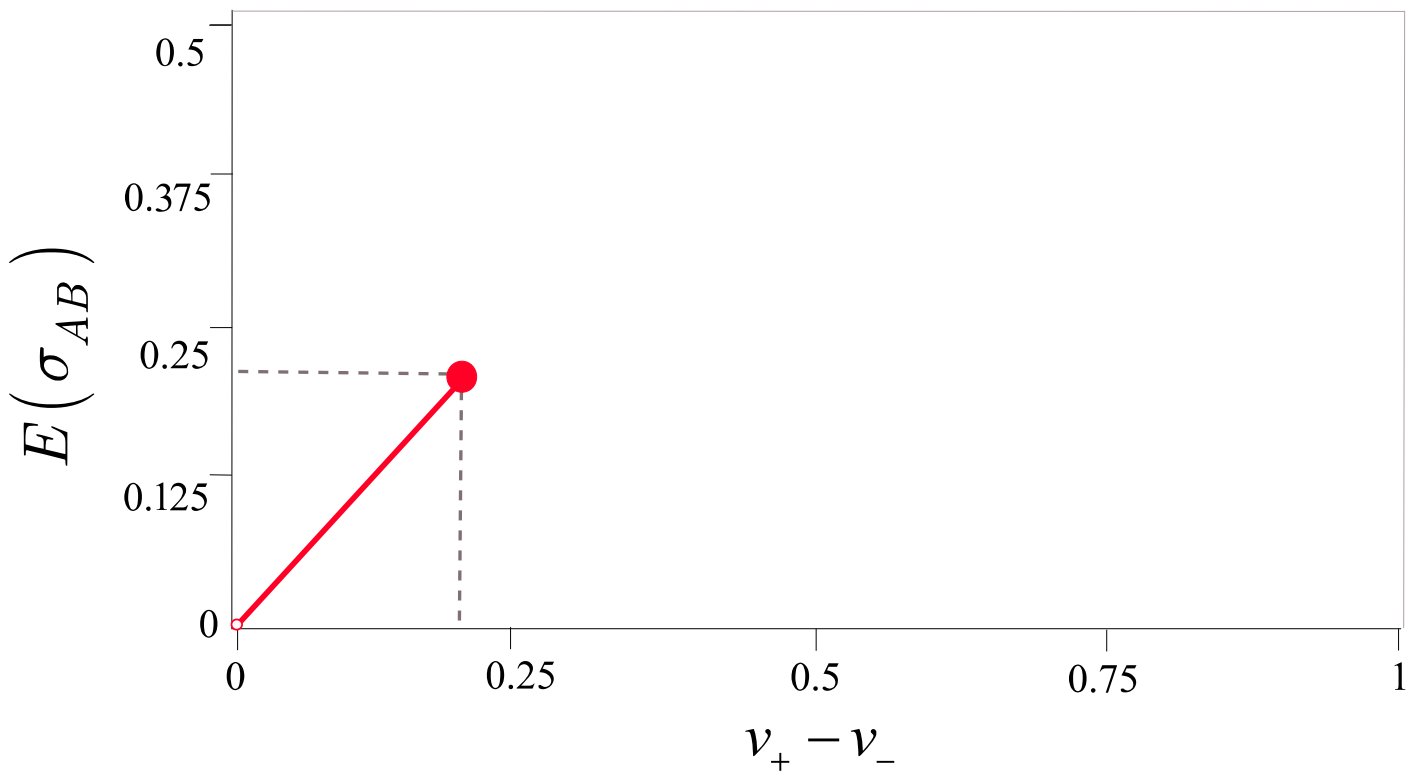

**Fig. S.2.** The amount of noise-generated distillable entanglement in function of the difference of the eigenvalues of channel output density matrix.

These results prove the statements of Theorem 3.

■

# S.2 Illustration of CC-property

In Fig. S.3, the $E\left(\cdot\right)$ relative entropy of entanglement in $\sigma_{AB}$ *in* function of the noise parameter $p$ of the first channel $\mathcal{N}_1$ is shown. To illustrate the effect of the noise of channel $\mathcal{N}_1$ on the amount of generated distillable entanglement, we characterized the Bell diagonal input (see (S.18)) system $\rho_{AB}$ as:

$$c_1 = \left(v_+ - v_-\right)_{in} = \frac{1}{3}, \tag{S.56}$$

$$c_2 = -\left(v_+ - v_-\right)_{in} = -\frac{1}{3} \tag{S.57}$$

and

$$c_3 = 1 - 2\cdot\left(v_+ - v_-\right)_{in} = 1 + 2\cdot c_2. \tag{S.58}$$

One can check readily that this input system is the same system given by formulas of (S.18) and (S.21), assuming $\left(v_+ - v_-\right)_{in} = \frac{1}{3}$. This system is separable, since $\left|c_1\right| + \left|c_2\right| + \left|c_3\right| \leq 1$, and $\max\left\{v_+, v_-, u_+, u_-\right\} \leq \frac{1}{2}$, where $v_+ = \frac{1}{2}$ and $v_- = \frac{1}{6}$.

The $p \geq \frac{1}{3}$ error probability of the phase flip channel $\mathcal{N}_1$ results in the decreasing amount of entanglement $E\left(\sigma_{AB}\right)$, for the increasing error probability $p$.

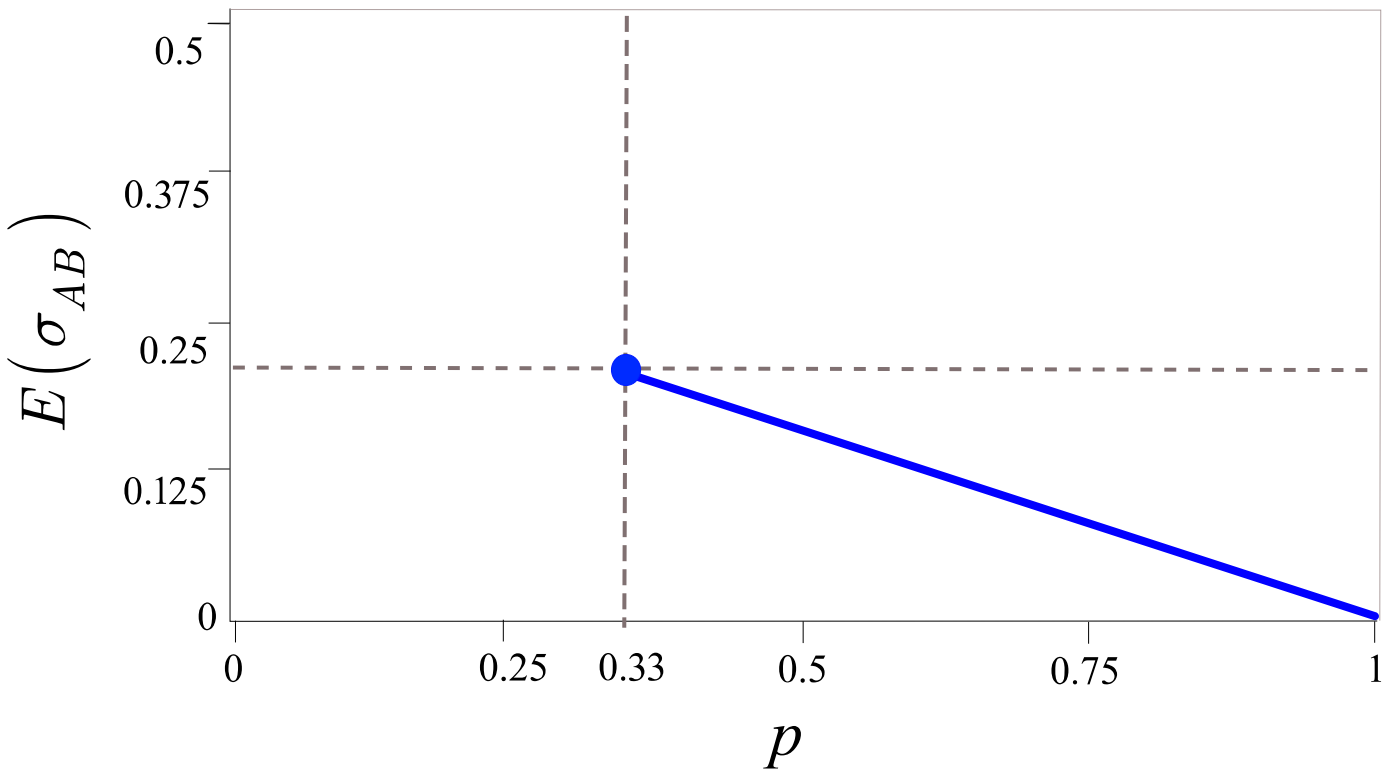

**Fig. S.3.** The amount of noise-generated distillable entanglement, assuming a phase flip channel $\mathcal{N}_1$ with $p \geq \frac{1}{3}$, an entanglement-breaking channel $\mathcal{N}_2$, and $c_1 = \frac{1}{3}, c_2 = -\frac{1}{3}$ and $c_3 = 1 + 2\cdot c_2$. The maximal entanglement $E\left(\sigma_{AB}\right) = \left(1-p\right)\cdot\left(v_+ - v_-\right)_{in} = \frac{(1-p)}{3} = \frac{2}{9}$ is obtained for $p = 1/3$.

For the given input system $\rho_{AB}$, the maximized amount of noise-generated entanglement over the channels $\mathcal{N}_1$ and $\mathcal{N}_2$ is

$$\begin{aligned} E\left(\sigma_{AB}\right) &= \min_{\rho_{AB}} D\left(\sigma_{AB} \| \rho_{AB}\right) = \max_{v_+ - v_-}\left(v_+ - v_-\right) \\ &= \left(1-p\right)\cdot\left(v_+ - v_-\right)_{in} = \frac{\left(1-p\right)}{3} = \frac{2}{9}, \end{aligned} \tag{S.59}$$

since for $p = 1/3$,

$$v_+ = \frac{1}{2} \text{ and } v_- = \frac{1}{4}\left(1 - \left(1 - \frac{2}{3}\cdot\left(1-p\right)\right) - \sqrt{\left(\left(1-p\right)\frac{1}{3} - \left(1-p\right)\frac{1}{3}\right)^2}\right) = \frac{5}{18}. \tag{S.60}$$

## S.2.1 Correlation Measures and Quantum Capacity

In this section, we derive the various correlation measures [15-31] for the output system $\sigma_{AB}$. These correlation measures can help to analyze further the properties of the Correlation Conversion property.

**Quantum Mutual Information**

The $I(\sigma_{AB})$ quantum mutual information function measures the *total* (i.e., both classical and quantum) correlation in the joint channel output state $\sigma_{AB}$. The quantum mutual information function of $\sigma_{AB}$ can be expressed as follows [15]:

$$I(\sigma_{AB}) = S(\rho_A) + S(\sigma_B) - S(\sigma_{AB}). \tag{S.61}$$

Using the eigenvalues of $\sigma_{AB}$, $I(\sigma_{AB})$ can be rewritten as [15]:

$$I(\sigma_{AB}) = S(\rho_A) + S(\sigma_B) + u_+ \log_2 u_+ + u_- \log_2 u_- + v_+ \log_2 v_+ + v_- \log_2 v_-, \tag{S.62}$$

where $u_+, u_-, v_+, v_-$ are the eigenvalues of $\sigma_{AB}$ (defined in (S.24) and (S.25)), and

$$S(\rho_A) = 1 - \frac{1}{2}(1-r)\log_2(1-r) - \frac{1}{2}(1+r)\log_2(1+r), \tag{S.63}$$

$$S(\sigma_B) = 1 - \frac{1}{2}(1-s)\log_2(1-s) - \frac{1}{2}(1+s)\log_2(1+s). \tag{S.64}$$

**Classical Correlation**

The $\mathcal{C}(\sigma_{AB})$ classical correlation function measures the *purely classical* correlation in the joint state $\sigma_{AB}$. The amount of purely classical correlation $\mathcal{C}(\sigma_{AB})$ in $\sigma_{AB}$ can be expressed as follows [16-18]:

$$\begin{aligned} \mathcal{C}(\sigma_{AB}) &= S(\sigma_B) - \tilde{S}(B|A) \\ &= S(\sigma_B) - \min_{E_k} \sum_k pkS\left(\sigma_{B|k}\right), \end{aligned} \tag{S.65}$$

where $\sigma_{B|k} = \frac{\langle k|\rho_A \sigma_B|k\rangle}{\langle k|\rho_A k\rangle}$ is the post-measurement state of $\sigma_B$, the probability of result $k$ is $p_k = d_{q_k} \langle k|\rho_A k\rangle$, while $d$ is the dimension of system $\rho_A$, $q_k$ makes up a normalized probability distribution in the rank-one POVM elements $E_k = q_k |k\rangle\langle k|$ of the POVM measurement operator [15-16]. The following definition will be used to compute the classical correlation (a more detailed derivation can be found in [41]):

$$\mathcal{C}(\sigma_{AB}) = S(\rho_A) - \min\{f_1, f_2, f_3\}, \tag{S.66}$$

where the functions $f_1, f_2$ and $f_3$ are defined as [15], [41]:

$$
\begin{aligned}
f_1 = & -\frac{1}{4}\left(1+r+s+c_3\right)\log_2 \frac{1}{2\left(1+s\right)}\left(1+r+s+c_3\right) \\
& -\frac{1}{4}\left(1-r+s-c_3\right)\log_2 \frac{1}{2\left(1+s\right)}\left(1-r+s-c_3\right) \\
& -\frac{1}{4}\left(1+r-s-c_3\right)\log_2 \frac{1}{2\left(1+s\right)}\left(1+r-s-c_3\right) \\
& -\frac{1}{4}\left(1-r-s+c_3\right)\log_2 \frac{1}{2\left(1+s\right)}\left(1-r-s+c_3\right),
\end{aligned} \tag{S.67}
$$

$$
f_2 = 1-\frac{1}{2}\left(1-\sqrt{r+c_1^2}\right)\log_2\left(1-\sqrt{r+c_1^2}\right)-\frac{1}{2}\left(1+\sqrt{r+c_1^2}\right)\log_2\left(1+\sqrt{r+c_1^2}\right), \tag{S.68}
$$

and

$$
f_3 = 1-\frac{1}{2}\left(1-\sqrt{r+c_2^2}\right)\log_2\left(1-\sqrt{r+c_2^2}\right)-\frac{1}{2}\left(1+\sqrt{r+c_2^2}\right)\log_2\left(1+\sqrt{r+c_2^2}\right). \tag{S.69}
$$

We note that Eqs. (S.66-S.69) are equivalent to Eq. (5) of [41] by performing local unitary transformations on the bipartite quantum state $\sigma_{AB}$.

**Quantum Discord**

The $\mathcal{D}\left(\sigma_{AB}\right)$ quantum discord function measures the *purely quantum* correlation in the joint state $\sigma_{AB}$. It is important to emphasize that this correlation measure does not identify the amount of distillable entanglement in the joint system $\sigma_{AB}$, hence it cannot be used to characterize the entanglement that generated by the channel. From the amount of quantum mutual information $I\left(\sigma_{AB}\right)$ and the classical correlation $\mathcal{C}\left(\sigma_{AB}\right)$ of output system $\sigma_{AB}$, the $\mathcal{D}\left(\sigma_{AB}\right)$ quantum discord can be expressed as

$$
\mathcal{D}\left(\sigma_{AB}\right) = I\left(\sigma_{AB}\right) - \mathcal{C}\left(\sigma_{AB}\right). \tag{S.70}
$$

Based on the previously-shown results, for the given channel output representation it can be rewritten in the following form:

$$
\begin{aligned}
& \mathcal{D}\left(\sigma_{AB}\right) = I\left(\sigma_{AB}\right) - \mathcal{C}\left(\sigma_{AB}\right) \\
& = S\left(\rho_A\right) + S\left(\sigma_B\right) + u_+ \log_2 u_+ + u_- \log_2 u_- + v_+ \log_2 v_+ + v_- \log_2 v_- \\
& -\left(S\left(\rho_A\right) - \min\left\{f_1, f_2, f_3\right\}\right) \\
& = S\left(\sigma_B\right) + u_+ \log_2 u_+ + u_- \log_2 u_- + v_+ \log_2 v_+ + v_- \log_2 v_- + \min\left\{f_1, f_2, f_3\right\}.
\end{aligned} \tag{S.71}
$$

**Quantum Coherent Information**

From the quantum discord $\mathcal{D}\left(\sigma_{AB}\right)$ and the classical correlation $\mathcal{C}\left(\sigma_{AB}\right)$ functions, the $I_{coh}\left(\sigma_{AB}\right)$ quantum coherent information of $\sigma_{AB}$ can be expressed as follows:

$$
\begin{aligned}
& I_{coh}\left(\sigma_{AB}\right) = \mathcal{D}\left(\sigma_{AB}\right) + \mathcal{C}\left(\sigma_{AB}\right) - 1 = \\
& = I\left(\sigma_{AB}\right) - \mathcal{C}\left(\sigma_{AB}\right) + \mathcal{C}\left(\sigma_{AB}\right) - 1 \\
& = I\left(\sigma_{AB}\right) - 1.
\end{aligned} \tag{S.72}
$$

Using the previously-derived results, it can also be rewritten as:

$$\begin{aligned}
& I_{coh}\left(\sigma_{AB}\right) \\
& = I\left(\sigma_{AB}\right) - 1 \\
& = S\left(\rho_A\right) + S\left(\sigma_B\right) + u_+ \log_2 u_+ + u_- \log_2 u_- + v_+ \log_2 v_+ + v_- \log_2 v_- - 1.
\end{aligned} \tag{S.73}$$

**Quantum Capacity**

The $Q\left(\mathcal{N}_1 \otimes \mathcal{N}_2\right)$ of the joint structure can be given as the maximization of the quantum coherent information $I_{coh}\left(\sigma_{AB}\right)$ of channel output system $\sigma_{AB}$,

$$\begin{aligned}
Q\left(\mathcal{N}_1 \otimes \mathcal{N}_2\right) &= \lim_{n\to\infty} \frac{1}{n} \max_{\forall \rho_A \rho_B} I_{coh}\left(\sigma_{AB}\right) \\
&= \lim_{n\to\infty} \frac{1}{n} \max_{\forall \rho_A \rho_B} \left(\mathcal{D}\left(\sigma_{AB}\right) + \mathcal{C}\left(\sigma_{AB}\right) - 1\right) \\
&= \lim_{n\to\infty} \frac{1}{n} \max_{\forall \rho_A \rho_B} \left(I\left(\sigma_{AB}\right) - 1\right).
\end{aligned} \tag{S.74}$$

From the previously-shown results it also can be expressed as follows:

$$\begin{aligned}
Q\left(\mathcal{N}_1 \otimes \mathcal{N}_2\right) &= \lim_{n\to\infty} \frac{1}{n} \max_{\forall \rho_A \rho_B} I_{coh}\left(\sigma_{AB}\right) \\
&= \lim_{n\to\infty} \frac{1}{n} \max_{\forall \rho_A \rho_B} \begin{pmatrix} S\left(\rho_A\right) + S\left(\sigma_B\right) + u_+ \log_2 u_+ + u_- \log_2 u_- + \\ v_+ \log_2 v_+ + v_- \log_2 v_- - 1 \end{pmatrix}.
\end{aligned} \tag{S.75}$$

From the previously-shown consequences, the following connection can be derived:

$$Q\left(\mathcal{N}_1 \otimes \mathcal{N}_2\right) = \lim_{n\to\infty} \frac{1}{n} \max_{\forall \rho_A \rho_B} \begin{pmatrix} S\left(\rho_A\right) + S\left(\sigma_B\right) + u_+ \log_2 u_+ + u_- \log_2 u_- \\ +\left(E\left(\sigma_{AB}\right) + v_-\right) \log_2 \left(E\left(\sigma_{AB}\right) + v_-\right) \\ +\left(v_+ - E\left(\sigma_{AB}\right)\right) \log_2 \left(v_+ - E\left(\sigma_{AB}\right)\right) - 1 \end{pmatrix}, \tag{S.76}$$

where $0 < E\left(\sigma_{AB}\right) \leq \frac{2}{9}$ and $u_+, u_-, v_+, v_-$ are non-negative real numbers. Assuming a Bell diagonal channel output state with $r = s = 0$, thus $S\left(\rho_A\right) = S\left(\sigma_B\right) = 1$, $Q\left(\mathcal{N}_1 \otimes \mathcal{N}_2\right)$ reduces to

$$\begin{aligned}
Q\left(\mathcal{N}_1 \otimes \mathcal{N}_2\right) &= \lim_{n\to\infty} \frac{1}{n} \max_{\forall \rho_A \rho_B} \begin{pmatrix} 1 + u_+ \log_2 u_+ + u_- \log_2 u_- \\ +\left(E\left(\sigma_{AB}\right) + v_-\right) \log_2 \left(E\left(\sigma_{AB}\right) + v_-\right) \\ +\left(v_+ - E\left(\sigma_{AB}\right)\right) \log_2 \left(v_+ - E\left(\sigma_{AB}\right)\right) \end{pmatrix} \\
&= \lim_{n\to\infty} \frac{1}{n} \max_{\forall \rho_A \rho_B} \left(1 - S\left(\sigma_{AB}\right)\right).
\end{aligned} \tag{S.77}$$

# Channel Output

Assuming the previously-characterized classically correlated input system $\rho_{AB}$ with $c_1 = \frac{1}{3}, c_2 = -\frac{1}{3}$ and $c_3 = 1 + 2 \cdot c_2$, channel $\mathcal{N}_1$ with error probability $p \geq \frac{1}{3}$, and the entanglement-breaking channel $\mathcal{N}_2$ the previously introduced correlation measures $I\left(\sigma_{AB}\right)$, $\mathcal{C}\left(\sigma_{AB}\right)$, $\mathcal{D}\left(\sigma_{AB}\right)$, $I_{coh}\left(\sigma_{AB}\right)$ and the amount of noise-generated entanglement $E\left(\sigma_{AB}\right)$ are compared in Fig. S.4. The results are shown for the composite system $AB$, where system $B$ is affected by the noise of $\mathcal{N}_1$.

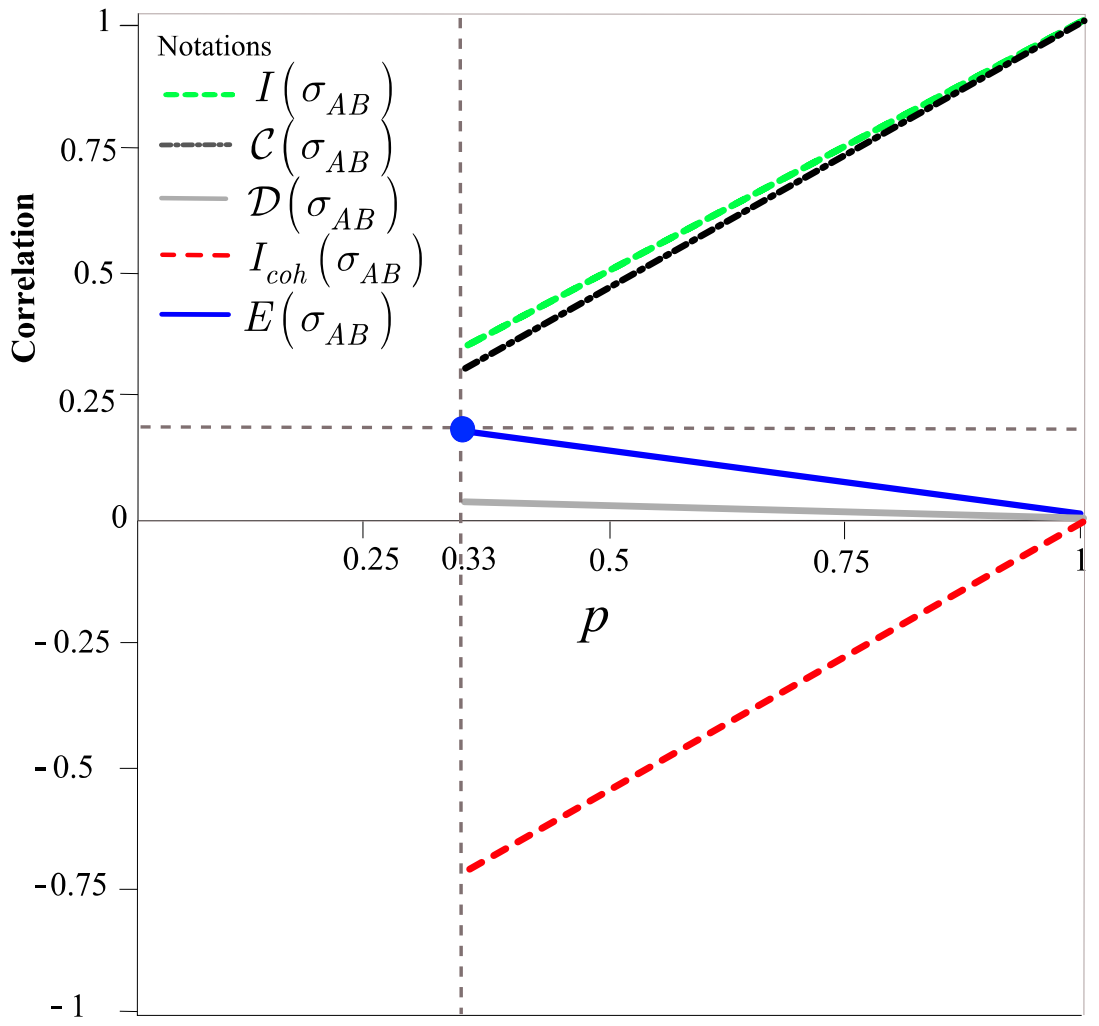

**Fig. S.4.** The amount of total correlation $I(\sigma_{AB})$, purely classical $\mathcal{C}(\sigma_{AB})$, purely quantum correlations $\mathcal{D}(\sigma_{AB})$, quantum coherent information $I_{coh}(\sigma_{AB})$ and the relative entropy of entanglement $E(\sigma_{AB})$ in the channel output system $\sigma_{AB}$, assuming $c_1 = \frac{1}{3}, c_2 = -\frac{1}{3}$ and $c_3 = 1 + 2 \cdot c_2$, with phase flip channel $\mathcal{N}_1$, $p \geq \frac{1}{3}$, and entanglement-breaking channel $\mathcal{N}_2$.

The coherent information $I_{coh}(\sigma_{AB})$, quantum discord $\mathcal{D}(\sigma_{AB})$ and the quantum entanglement are *quantum correlations. The purely classical correlation is measured by* $\mathcal{C}(\sigma_{AB})$. The quantum mutual information $I(\sigma_{AB})$ measures both classical and quantum correlations. From these correlations, the quantum entanglement can be achieved in $\sigma_{AB}$ if only the measurement $\{\Lambda_C\}$ of the entanglement-breaking channel $\mathcal{N}_2$ on $C$ has been resulted in 0. If $\mathcal{N}_2$ measured 1, then only *classical correlations* will be available in $\sigma_{AB}$. Increasing $p$ of $\mathcal{N}_1$, the total correlation $I(\sigma_{AB})$ and classical correlation $\mathcal{C}(\sigma_{AB})$ start to increase, while the discord $\mathcal{D}(\sigma_{AB})$ and the coherent information $I_{coh}(\sigma_{AB})$ start to decrease. At $p = 1$, $I(\sigma_{AB})$ reduces to $\mathcal{C}(\sigma_{AB})$, and $\mathcal{D}(\sigma_{AB})$ to 0, while the $I_{coh}(\sigma_{AB})$ coherent information will be $I(\sigma_{AB}) - 1 = \mathcal{C}(\sigma_{AB}) - 1 = 0$, along with $E(\sigma_{AB}) = 0$, hence the noise of the channel destroys every quantum correlations in the channel output system $\sigma_{AB}$.

# S.3 Application in Quantum Repeater Networks

The Correlation Conversion scheme can directly be applied in long-distance quantum communication or in quantum repeater networks. To share $l = \lfloor n(1-p) \rfloor$, $p \geq \frac{1}{3}$, entangled systems between Alice and Bob, Alice has to generate only $n$ instances of system $ABC$, then send the $n$ systems of $B$ over channel $\mathcal{N}_1$ and $n$ systems of $C$ over $\mathcal{N}_2$. The second channel generates $n$ classical bits to Bob.

As depicted in Fig. S.5, in Phase 1, Bob receives the $n$-length qubit string of $B$ from $\mathcal{N}_1$. In Phase 2, Bob receives an $n$-length classical bit string from $\mathcal{N}_2$ to identify the indices of the entangled states received in Phase 1. The indices of the classical bitstring determine unambiguously the indices of Bob's entangled states in $B$. If the $i$-th classical bit is 0, then Bob will know that the $i$-th state of $B$ is entangled. If its value is 1, then Bob will know that the given index of $B$ does not contain any entanglement.

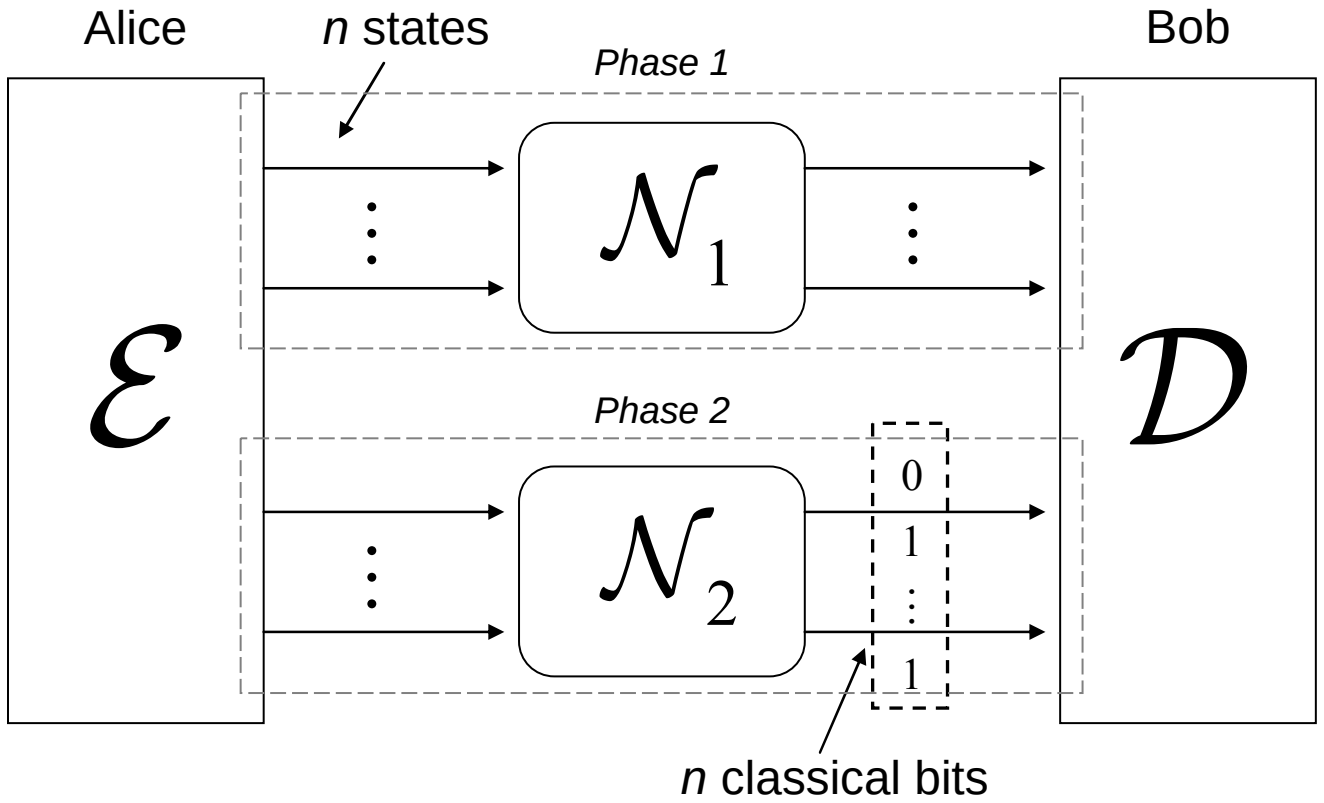

**Fig. S.5.** Correlation Conversion in a practical application. In *Phase* 1, Alice sends an $n$-length qubit string of $B$ to the first channel, and an $n$-length qubit string of $C$ in *Phase* 2 to the entanglement-breaking channel. Bob receives the states of $B$, then the classical bit string from the second channel. The indices of the classical bitstring identify unambiguously the indices of the entangled systems in the string.

The proposed process allows easy implementation, and from now on, practical entanglement sharing can be realized without the need of entanglement transmission. The development of quantum repeater networks will also be possible without the expensive process of entanglement transmission by using only standard noisy quantum channels, with significantly lower effort and development costs.

# S.4 Extension to Qudit Systems

The $d = 2$ dimensional results can be extended to higher dimensions, as follows. Using the notations of [9], the bipartitions of the tripartite state $ABC$ will be referred as $A : BC, B : AC$ and $C : AB$. The Schmidt decomposition of the $d$-dimensional input system $ABC$ can be given by $|\psi\rangle_{ABC} = \sum_i x_i |x_i\rangle_X |\tilde{x}_i\rangle_{\tilde{X}}$, where $x_i = a_i, b_i, c_i$ are the Schmidt coefficients, $X = A, B, C$ and $\tilde{X} = BC, AC, AB$, while $|\tilde{b}_i\rangle_{AC}$ are orthogonal vectors [9]. The action of the $d$-dimensional quantum channels $\mathcal{N}_1$ (phase flip) and $\mathcal{N}_2$ (entanglement-breaking channel) can be modeled as a unitary operation $U_{AC}$, which acts on the qudit subsystems $A$ and $C$.

One can also introduce parameter $\tau$, as follows:

$$\tau = \frac{1}{1+Md} = \left(v_+ - v_-\right)_{in}, \tag{S.78}$$

where $\left(v_+ - v_-\right)_{in}$ is the difference of the eigenvalues of the qudit subsystem $AB$, $M = \max\left(b_1 b_2, c_1 c_2\right)$, $M < a_1 a_2$, and $d$ is the dimension of the system [9].

Using (S.78), the input system is

$$\rho_{ABC} = \tau \left|\phi\right\rangle\left\langle\phi\right|_{AB} \otimes \left|0\right\rangle\left\langle 0\right|_C + \left(1-\tau\right)\frac{I}{d} I, \tag{S.79}$$

where $\frac{I}{d} I$ is the $d$-dimensional maximally mixed state and system $AB$ is

$$\left|\phi\right\rangle_{AB} = \sum_i b_i \left|i\right\rangle_A \left|b_i\right\rangle_B. \tag{S.80}$$

Applying the unitary $U_{AC}$ on (S.79), where $d_B \leq d_A$, and $U_{AC}\left|i\right\rangle_A\left|0\right\rangle_C = \left|\tilde{b}_i\right\rangle_{AC}$, results in the channel output qudit system

$$\sigma_{ABC} = \gamma\tau\left|\psi\right\rangle\left\langle\psi\right|_{ABC} + \left(1-\gamma\tau\right)\frac{I}{d} I, \tag{S.81}$$

where $\left|\phi\right\rangle\left\langle\phi\right|_{AB} \neq \left|\psi\right\rangle\left\langle\psi\right|_{AB}$, $\gamma = \left(1-p\right)$ and

$$\gamma\tau = \left(1-p\right)\cdot\left(v_+ - v_-\right)_{in} < \left(v_+ - v_-\right)_{in}, \tag{S.82}$$

while $p$ is the noise of the $d$-dimensional channel $\mathcal{N}_1$, and $\left|\psi\right\rangle_{ABC}$ is a pure tripartite state. The initial system can be expressed as

$$\rho_{ABC} = U_{AC}^{\dagger}\sigma_{ABC}U_{AC}. \tag{S.83}$$

Sending the flag $C$ into $\mathcal{N}_2$, the remaining $AB$ subsystem can be rewritten as

$$\sigma_{AB} = \tau\gamma\left|\psi\right\rangle\left\langle\psi\right|_{AB} + \left(1-\tau\gamma\right)\frac{I}{d} I. \tag{S.84}$$

The state in (S.84) is *entangled* if only if $\tau\gamma > \frac{1}{1+a_1 a_2 d}$, where $a_1$, $a_2$ are the two largest Schmidt coefficients of the bipartite state $\left|\psi\right\rangle_{AB}$. Since the required value of $\tau\gamma$ is lower bounded by the Schmidt coefficients, it shows that there exists a bipartite system $\left|\psi\right\rangle_{AB}$ and error probability $p$ of $\mathcal{N}_1$, for which the quantity $\frac{1}{1+a_1 a_2 d}$ picks up its *minimum* (i.e., the Schmidt coefficients $a_1$, $a_2$ are *maximal* [9]) from a finite range. In this case, the qudit channel output state $\sigma_{AB}$ becomes only $A:BC$-entangled, while across the bipartitions $B:AC, C:AB$ the qudit state remain separable, hence the CC-property can be applied for qudit inputs and channels.

In future work our aim is to find other possible, well-defined channel combinations to further demonstrate the potential of CC-property.

# References

[1] W. J. Munro, K. A. Harrison, A. M. Stephens, S. J. Devitt, and K. Nemoto, Nature Photonics, 10.1038/nphoton.2010.213, (2010)

[2] L. Hanzo, H. Haas, S. Imre, D. O'Brien, M. Rupp, L. Gyongyosi, Wireless Myths, Realities, and Futures: From 3G/4G to Optical and Quantum Wireless, *Proceedings of the IEEE, Volume: 100, Issue: Special Centennial Issue,* pp. 1853-1888. (2012).

[3] S. Imre and L. Gyongyosi, *Advanced Quantum Communications - An Engineering Approach*, Publisher: Wiley-IEEE Press (New Jersey, USA), John Wiley & Sons, Inc., The Institute of Electrical and Electronics Engineers. Hardcover: 524 pages, ISBN-10: 1118002369, ISBN-13: 978-11180023, (2012).

[4] R.V. Meter, T. D. Ladd, W.J. Munro, K. Nemoto, System Design for a Long-Line Quantum Repeater, IEEE/ACM Transactions on Networking 17(3), 1002-1013, (2009).

[5] C. H. Bennett and G. Brassard. Quantum cryptography: Public key distribution and coin tossing. In Proc. IEEE International Conference on Computers, Systems, and Signal Processing, pages 175–179, (1984).

[6] K. G. Paterson, F. Piper, and R. Schack. Why quantum cryptography? http://arxiv.org/quant-ph/0406147, (2004).

[7] C. Elliott, D. Pearson, and Gregory Troxel. Quantum cryptography in practice. In Proc. SIGCOMM 2003. ACM, (2003).

[8] T. S. Cubitt, F. Verstraete, W. Dür, and J. I. Cirac, Phys. Rev. Lett. 91, 037902 (2003).

[9] T. K. Chuan, J. Maillard, K. Modi, T. Paterek, M. Paternostro, and M. Piani, Quantum discord bounds the amount of distributed entanglement, arXiv:1203.1268v3, Phys. Rev. Lett. 109, 070501 (2012).

[10] A. Streltsov, H. Kampermann, D.r Bruß, Quantum cost for sending entanglement, arXiv:1203.1264v22012. Phys. Rev. Lett. 108, 250501 (2012)

[11] A. Kay, Resources for Entanglement Distribution via the Transmission of Separable States, arXiv:1204.0366v4, Phys. Rev. Lett. 109, 080503 (2012).

[12] J. Park, S. Lee, Separable states to distribute entanglement, arXiv:1012.5162v2, Int. J. Theor. Phys. 51 (2012) 1100-1110 (2010).

[13] N. J. Cerf, Quantum cloning and the capacity of the Pauli channel, arXiv:quant-ph/9803058v2, Phys.Rev.Lett. 84 4497 (2000).

[14] J. Maziero, L. C. Celeri, R. M. Serra, V. Vedral, Classical and quantum correlations under decoherence, arXiv:0905.3396v3 (2009).

[15] B. Li, Z. Wang, S. Fei, Quantum Discord and Geometry for a Class of Two-qubit States, arXiv:1104.1843v1, (2011).

[16] M. D. Lang, and C.M. Caves, Phys. Rev. Lett. 105, 150501 (2010).

[17] F. Hui-Juan, L. Jun-Gang, Z, Jian, and S. Bin. Connections of Coherent Information, Quantum Discord, and Entanglement, Commun. Theor. Phys. 57, 589–594. (2012).

[18] M. Ali, A.R.P. Rau, and G. Alber, Phys. Rev. A 82, 069902 (2010).

[19] L. Mazzola, J. Piilo, and S. Maniscalco, Phys. Rev. Lett. 104, 200401 (2010).

[20] B. Bylicka and D. Chruscinski, Phys. Rev. A 81, 062102 (2010).

[21] T. Werlang, S. Souza, F.F. Fanchini, and C.J. Villas Boas, Phys. Rev. A 80, 024103 (2009).

[22] M. S. Sarandy, Phys. Rev. A 80, 022108 (2009).

[23] A. Ferraro, L. Aolita, D. Cavalcanti, F. M. Cucchietti, and A. Acin, Phys. Rev. A 81, 052318 (2010).

[24] F. F. Fanchini, T. Werlang, C.A. Brasil, L.G.E. Arruda, and A.O. Caldeira, Phys. Rev. A 81, 052107 (2010).

[25] B. Dakic, V. Vedral, and C. Brukner, Phys. Rev. Lett. 105, 190502 (2010).

[26] K. Modi, T. Paterek, W. Son, V. Vedral, and M. Williamson, Phys. Rev. Lett. 104, 080501 (2010).

[27] N. Li and S. Luo, Phys. Rev. A 76, 032327 (2007);

[28] S. Luo, Phys. Rev. A 77, 042303 (2008).

[29] C. Q. Pang, F. Zhang, Y. Jiang, M. Liang, J. Chen, Most robust and fragile two-qubit entangled states under depolarizing channels, arXiv:1202.2798, Quantum Information and Computation, Vol. 13, No. 7&8, 0645–0660, (2013).

[30] T. Konrad, F. De Melo, M. Tiersch, C. Kasztelan, A. Aragao, and A. Buchleitner, Nature physics 4, 99 (2007).

[31] A. Peres, Separability Criterion for Density Matrices, Phys. Rev. Lett. 77, 1413–1415 (1996)

[32] M. Horodecki, P. Horodecki, R. Horodecki, Separability of Mixed States: Necessary and Sufficient Conditions, Phys. Lett. A 223, 1-8 (1996)

[33] K. Życzkowski and I. Bengtsson, Geometry of Quantum States, Cambridge University Press, (2006).

[34] S. L. Woronowicz, Positive maps of low dimensional matrix algebras, Rep. Math. Phys. 10, 165–183. (1976)

[35] W. Dür, J. I. Cirac, and R. Tarrach, Phys. Rev. Lett. 83, 3562 (1999)

[36] W. Dür and J. I. Cirac, Phys. Rev. A 61, 042314 (2000).

[37] K. Bradler, P. Hayden, D. Touchette, M. M. Wilde, Trade-off capacities of the quantum Hadamard channels, Physical Review A 81, 062312 (2010).

[38] C. H. Bennett et al., PRA 54, 3824 (1996)

[39] F-L. Zhang, Y. Jiang, M-L. Liang, Speed of disentanglement in multi-qubit systems under depolarizing channel, Annals of Physics (2013).

[40] P. Horodecki, M. Horodecki, and R. Horodecki, "Binding entanglement channels," J. Mod. Opt., vol. 47, pp. 347–354, 2000.

[41] Y. Huang, Quantum discord for two-qubit X states: Analytical formula with very small worst-case error, Phys. Rev. A 88, 014302 (2013).

[42] Laszlo Gyongyosi, Sandor Imre: Distillable Entanglement from Classical Correlation, Proceedings of SPIE Quantum Information and Computation XI, 2013.